\documentclass[12pt,
superscriptaddress,
 amsmath,amssymb,
 aps,
floatfix
]{revtex4-2}
\usepackage[utf8]{inputenc}
\usepackage[caption=false]{subfig}
\usepackage[version=4]{mhchem}
\usepackage{siunitx}
\usepackage{xcolor}

\usepackage{listings}
\usepackage{epstopdf, epsfig}
\usepackage{graphicx}
\usepackage{dcolumn}
\usepackage{bm}

\begin{document}

\preprint{APS/123-QED}

\title{On the biogenic hydrodynamic transport of upward and downward cruising copepods}

\author{Yunxing Su}\thanks{These authors contributed equally to this work.}
\affiliation{Mechanical and Industrial Engineering Department, University of Minnesota Duluth, Duluth, MN 55812, USA} 

\author{Rui Zhu}\thanks{These authors contributed equally to this work.}
\affiliation{Institute of Urban and Industrial Water Management, Technische Universität Dresden, Dresden 01062, Germany}

\author{Eckart Meiburg}
\affiliation{Department of Mechanical Engineering, University of California, Santa Barbara, Santa Barbara, CA 93106-5070, USA}

\author{Monica M. Wilhelmus}\email[]{Corresponding author: mmwilhelmus@brown.edu}
\affiliation{Center for Fluid Mechanics, School of Engineering, Brown University, Providence RI 02912 USA}

\date{\today}


\maketitle

\section*{Abstract}
Mesozooplankton aggregations undergoing vertical migrations in the upper ocean have been hypothesized to have an important role in the redistribution of carbon, nutrients, and oxygen via biogenic hydrodynamic transport (BHT). While laboratory studies have demonstrated how swarm-induced hydrodynamic instabilities can drive large-scale transport in strongly stratified environments, measurements are usually performed with model organisms that differ in morphology and swimming mode from ecologically relevant marine species. To bridge this gap, we conducted experiments with copepods and analyzed upward and downward trajectories to identify differences in flow fields, force distribution, and BHT for these two scenarios. Using two-dimensional bright-field Particle Image Velocimetry (PIV), we quantified the near-body velocity field and found that the average downward swimming speed significantly exceeds the average upward swimming speed, with the flow fields exhibiting direction-dependent characteristics. We incorporated these findings into a continuum squirmer model to estimate the swimmer-induced drift volume and mixing efficiency, focusing on the effects of the reduced gravity of the swimmers and the density stratification of the surrounding fluid. Our simulations reveal that both the excess weight of the organisms and the fluid stratification strongly constrain the net BHT. This study provides a critical step toward integrating lab-based models of marine mesozooplankton with remote sensing data to incorporate vertical migrations into global ocean models with realistic biogeochemistry and assess their ecological significance in actively sustaining local ecosystems.
\\

{\bf Keywords}: {copepod, drift volume, biogenic hydrodynamic transport, squirmer model, particle image velocimetry}

\section{Introduction}
Mesozooplankton are abundant and widely distributed across diverse marine environments, from coastal shelves to the open ocean. Many species are well known for undergoing diel vertical migrations (DVM) \cite{dewar2006does,bianchi2013intensification}--synchronized, large-scale population shifts that are thought to be largely driven by the need for predator avoidance \citep{bandara2021two}. In the fluid dynamics literature, DVM has long been hypothesized to trigger hydrodynamic instabilities leading to large-scale transport (hereafter referred to as biogenic hydrodynamic transport or BHT) \citep{kunze2006observations}. Important insights have been gathered from controlled laboratory studies \citep{wilhelmus2014,houghton2018vertically,mohebbi2024measurements} and numerical simulations of this process \citep{jiang2002flow_II,thiffeault2010stirring,wang2015biogenic,ouillon2020active}, though some studies have reported small or negligible mixing rates \cite{visser2007biomixing,wagner2014mixing,noss2014direct,kunze2019biologically}. This highlights the need to better assess the mixing efficiency and to establish an accurate evaluation of the net effect of BHT in real marine ecosystems, where it is challenging to directly link lab-based or idealized simulation results to dynamically complex ocean environments. 

Building on the conceptual framework of BHT \cite{wilhelmus2019effect}, we focus on how the swimming characteristics of ecologically relevant organisms affect net transport and, consequently, irreversible mixing. As one of the most abundant multicellular taxa in the ocean, copepods are invariably a primary driver of BHT. Extensive research has characterized their behavioral traits \cite{yen1996advertisement,yen2000life,jiang2004hydrodynamics}---feeding, sensory perception, swarming, mating, and predator evasion---alongside their associated flow features. In these studies, flow characterization has involved a variety of methods, including laboratory experiments \citep{wickramarathna2014hydrodynamic,michalec2015turbulence}, theoretical models \citep{jiang2002flow_I}, and numerical simulations \citep{ardekani2017transport,borazjani2020numerical}. Key examples include the study by \citet{stamhuis1995quantitative}, which integrated Particle Image Velocimetry (PIV) and Particle Tracking Velocimetry (PTV) to capture the far-field flow of cruising copepods, while Jiang et al. \cite{jiang2002flow_I,jiang2002flow_II} employed theoretical models and numerical simulations to investigate the three-dimensional flow patterns around freely swimming copepods undergoing hovering, sinking, together with upward, backward and forward swimming. In their analysis, organism morphology is neglected and the motion of the swimmer is modeled as a translating sphere in Stokes flow. While the typical Reynolds number for a swimming copepod is $Re \sim 1$, inertia-free models still yield valuable insights, particularly regarding how the copepod-induced flow fields vary with different swimming modes---a variation governed by the interplay between active propulsion and the excess weight (the net difference between the buoyancy and the weight of the organism). This force distribution was further elucidated by \citet{malkiel2003three}, who modeled the swimming organism by integrating a Stokeslet model along with the measured flow fields of a freely swimming copepod to estimate the propulsive forces and the excess weight. To bridge the gap between these low-Reynolds-number approximations and inertial effects experienced by real organisms, squirmer models have since enabled three-dimensional simulations across both the Re $<<1$ and Re $\sim1$ regimes \citep{khair2014expansions,li2014effect,ouillon2020active}.

Beyond directly analyzing free-swimming behavior, insights regarding the hydrodynamic signature of copepods have also been gathered from lab studies using tethered organisms. For instance,  \citet{catton2007quantitative} observed noticeable differences in the streamline patterns and the associated vorticity, dissipation rate, and strain rate fields derived from two-dimensional PIV measurements of tethered and free-swimming {\it Euchaeta antarctica}. Similarly, \citet{noss2012zooplankton} observed that tethered \textit{Daphnia} generate significantly higher dissipation rates than their freely swimming counterparts. These discrepancies arise from the force imbalance inherent to tethered swimmers compared to their freely swimming counterparts, suggesting that the presence of excess weight could indeed have an important effect on the topology of local flow fields around self-propelled organisms \cite{strickler1982calanoid}. High-speed tomographic PIV have allowed further inspection of the three-dimensional complexities of near-body fields \cite{murphy2012high}, which seem to be affected by distinct biological strategies. For instance, \citet{kiorboe2014flow} found that mesozooplankton effectively lower the risk of detection by rheotactic predators by reducing flow disturbances when decoupling feeding from swimming.  Collectively, these findings underscore how the force distribution in a swimming organism can affect its hydrodynamic footprint, shifting the system away from idealized neutrally buoyant models.
 
Besides the inherent force distribution resulting from self-propulsion, external factors can play an important role in transport and mixing. Ambient fluid stratification, for instance, is relevant for BHT in marine environments \cite{ardekani2017transport}. Although not directly applicable to the active swimming scenario, direct numerical simulations of spherical particles sedimenting in linearly stratified fluids have illustrated the effects of buoyancy-driven return flows induced by the vertical displacement of fluid across isopycnals \cite{doostmohammadi2014numerical}. It is important to note that transport in this case may be more efficient than by any active swimming counterpart due to the slower spatial attenuation rate of the local velocity field in the case of sedimenting particles. Swimming-induced disturbances, though continuous, can be modeled as evolving vortex systems and thus their interaction with density gradients can also be thought of being tightly linked to the net mixing efficiency. In fact, \citet{olsthoorn2017three} investigated how vortex rings induce mixing in density-stratified fluids by examining their three-dimensional time-resolved velocity fields reconstructed from PIV measurements. \citet{su2023asymmetry} further pointed out that the production of baroclinic vorticity directly affects the penetration of a translating vortex ring in a two-layered system, whereby the scale and propagation of such vortices are coupled to their orientation relative to the density gradient. Given that the ocean is stratified, characterizing this asymmetry in vortex dynamics is essential for evaluating the net BHT generated by the massive, directional migrations of zooplankton aggregations.

The interplay between swimmer orientation and transport in stratified media remains experimentally challenging to address and has received considerably less attention than swimming in homogeneous fluids. Clear differences have been documented in the migration speeds of mesozooplankton during ascent versus descent, observed at both the individual \cite{tyrell2020copepod} and aggregation \cite{bianchi2016global} scales. Furthermore, numerical investigations using squirmer models at intermediate Reynolds numbers indicate that mixing efficiency is enhanced for both positively and negatively buoyant swimmers relative to neutrally buoyant counterparts \citep{wang2015biogenic}. This orientation-dependent behavior may be rooted in biomechanical adaptations; for instance, copepods have been shown to alter their swimming gait depending on their orientation relative to gravity \citep{tack2024ups}. Such asymmetries, arising from the coupling of gravitational effects and density stratification with individual swimming kinematics, may be amplified during large-scale migrations. Consequently, these effects likely play a pivotal role in BHT and the subsequent redistribution of dissolved oxygen, nutrients, and carbon within the global ocean.

In this study, we employ experiments and simulations to examine how the swimming direction relative to gravity influences speed and the near-field flow fields of copepods swimming in the low-to-intermediate inertial regime ($Re$ $\sim$ 0.25 to 0.5) within the broader $Re$ $\sim$ 1 category. The Darwinian drift volume and mixing efficiency are quantified to estimate the BHT associated with the studied organisms. The remaining sections of the paper are organized as follows. The experimental methods and numerical approach are introduced in Sections~\ref{exp method} and ~\ref{num approach}. Experimental data for individual copepod swimming speed and the corresponding induced two-dimensional flow fields are presented in Section~\ref{Experimetnal results}. Numerical simulations using the squirmer model based on our experimental data are shown in  Section~\ref{simulation results}, where the drift volume and mixing efficiency are assessed and discussed in both homogeneous and density stratified fluids. The paper closes with a discussion of the results and presents an outlook for future studies.

\section{Experimental methods}\label{exp method}
\subsection{Model organism}\label{sec:Mo}
\begin{figure}
    \centering
    \includegraphics[width = 1\textwidth]{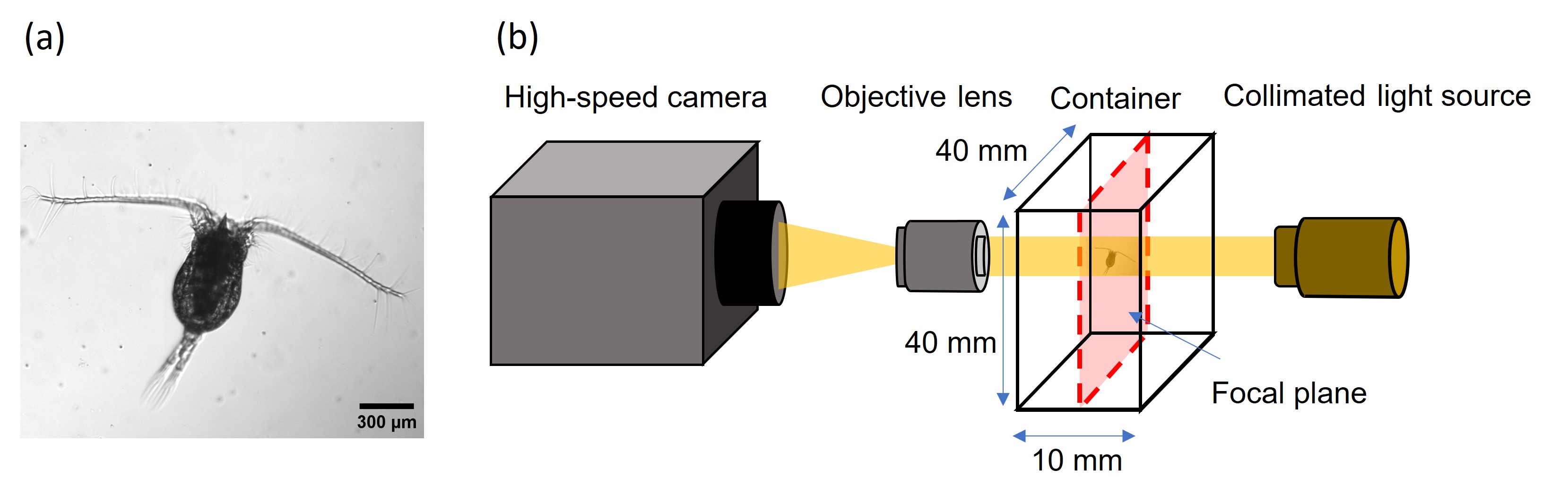}
    \caption{Model organism and experimental setup. (a) Adult copepods ({\it Parvocalanus crassirostris}). (b) Experimental setup (in-house micro-bright-field PIV setup) for individual copepod swimming speed and flow field measurements. The focal plane (red region) is vertical, allowing for measurements in the vertical direction. More details about the experimental setup can be obtained in \cite{gemmell2014new}.}
    \label{fig:setup}
\end{figure}

Lab-cultured copepods ({\it Parvocalanus crassirostris}, Fig.~\ref{fig:setup}(a)) were used as model organisms in the experiments (body length 0.53 $\pm$ 0.08 mm). The cruising speed (1.24 $\pm$ 1.08 mm s$^{-1}$) corresponds to a Reynolds number around 0.5. This copepod species is slightly negatively buoyant and is commonly found throughout the tropical and subtropical Atlantic Ocean, where it performs DVM. Copepods are known to be phototactic, meaning they respond to specific wavelengths and intensities of light, which is typical among mesozooplankton species undergoing DVM. Leveraging this characteristic, we designed an experiment to create and control a vertically migrating copepod swarm. Note that even though adults were used for the experiment, a small amount of nauplii were in the solution.

\subsection{Experimental setup}
Our experimental PIV setup is similar to configurations presented in the literature (e.g., \cite{khodaparast2013micro,gemmell2014new,herrera2021spatiotemporal}). Here, we used an in-house microscope equipped with a high-speed camera (Photron R2, 2048 pixels x 2048 pixels), an objective lens (M5, Newport Optics), an acrylic tank (4 cm $\times$ 1 cm $\times$ 4 cm), and a collimated light source (MI-152, Dolan-Jenner), as presented in Fig.~\ref{fig:setup}(b). We oriented the focal plane vertically to enable image capture while the copepods were cruising vertically. The resulting field of view is approximately 2 mm by side.

Time-resolved PIV measurements were acquired by filming complete swimming cycles of the self-propelled copepods at 1,440 fps. The fluid was pre-seeded with biodegradable tracer particles (algae, {\it Nannochloropsis}, $\sim$ 2 $\mu$m in diameter), which have been shown to have minimal effects on organisms \citep{gemmell2014new,Biodegradable2024}.  The velocity fields were computed in DaVis 10 (LaVision) following standard techniques \citep{wilhelmus2014,di2020bioinspired,ribeiro2021wake}. This involved using a multi-pass cross-correlation method with decreasing interrogation window size (initial at 96 pixels $\times$ 96 pixels, final at 48 pixels $\times$ 48 pixels) with a 50\% overlap \citep{zhu2020nonlinear, Santos2023pleobot}. Finally, we used the Matlab tool (DLT8a) \citep{hedrick2008software} to track the position of all the copepods cruising within the field of view. The instantaneous positions were used to calculate the average swimming speeds of the steadily cruising organisms.

\section{Simulation approach}\label{num approach}
We simulated the vertical migration of individual free-swimming copepods in a stratified water column by combining the squirmer model with fully coupled Direct Numerical Simulations (DNS) of the velocity and density fields of the surrounding fluid.

\subsection{Governing equations}
As introduced by \citet{lighthill1952squirming}, the squirmer model was adapted by \citet{blake1971spherical} and was traditionally used for swimmers in Stokes flows. It was recently extended to finite Reynolds number flows \citep{chisholm2016squirmer,ouillon2020active}. In this model, a swimmer generates thrust by forcing the fluid radially and tangentially along the surface of its body. Here, we consider a previously validated reduced-order version that effectively neglects the radial velocity and only considers the first two modes of the squirming motion, thereby simplifying the equations of organism locomotion \citep{chisholm2016squirmer}. With this simplification,  the magnitude of the tangential velocity in the reference frame of the moving object is given by
\begin{equation}
u_{\theta} = B_1\mathrm{sin}\theta+B_2\mathrm{sin}\theta\mathrm{cos}\theta\;,
\label{TV_t}
\end{equation}
where $\theta$ is the polar angle between the normal vector of $u_{\theta}$ and the swimming direction, $B_1$ and $B_2$ are the first and second modes, respectively, whereby $B_1$ is responsible for propulsion and $B_2$ determines the intensity of the stresslet exerted by the swimmer \citep{ouillon2020active}. The ratio $\beta=B_2/B_1$ indicates the squirming mode, with $\beta>0$ defining a puller generating thrust in front of the body and $\beta<0$ defining a pusher generating thrust behind the body.

In this study, we simulate the vertical migration of squirmer swimmers in a stratified water column by imposing the relative squirming velocity at spherical Lagrangian tracer points within the DNS setup. This is performed by solving an ordinary differential equation for the translational velocity of the tracer particles $\boldsymbol u_p=(u_p,v_p,w_p)^T$,
\begin{equation}
m_p\frac{\mathrm{d}\boldsymbol u_p}{\mathrm{d}t}=\oint_{\Gamma_p}\boldsymbol\tau\boldsymbol\cdot\boldsymbol n\ \mathrm{d}A+V_p(\rho_p-\rho_f)\boldsymbol g,
\label{TV}
\end{equation}
and its angular velocity $\boldsymbol\omega_p=(\omega_{p,x},\omega_{p,y},\omega_{p,z})^T$,
\begin{equation}
I_p\frac{\mathrm{d}\boldsymbol \omega_p}{\mathrm{d}t}=\oint_{\Gamma_p}\boldsymbol r\times(\boldsymbol\tau\boldsymbol\cdot\boldsymbol n)\ \mathrm{d}A,
\label{AV}
\end{equation}
where $m_p$ is the particle mass, $I_p=\pi\rho_pd^5_p/60$ is the moment of inertia, and $\rho_p$, $d_p$ and $V_p$ are the particle density, diameter and volume, respectively. Here, $\Gamma_p$ is the fluid-particle interface and $\boldsymbol\tau$ the hydrodynamic stress tensor. The vector $\boldsymbol n$ denotes the outward-pointing normal on the interface $\Gamma_p$, and $\boldsymbol r$ indicates the position vector from the center of the particle to a point on its surface. We only simulate individual swimmers and thus there are no forces due to collisions. For more details of the fluid solver and its validation, we refer the reader to \citet{biegert2017collision}. 

Considering Eqs. 1 - 3 above, 
\begin{equation}
\boldsymbol u = \boldsymbol u_p + \boldsymbol\omega_p \times \boldsymbol r + u_{\theta}\boldsymbol e_{\theta}\ \mathrm{on}\ \ \Gamma_p,
\label{flv}
\end{equation}
where $u_{\theta}\boldsymbol e_{\theta}$ is the squirmer slip velocity imposed on the particle surface, acting along the direction of the unit vector $\boldsymbol e_{\theta}$. 

The near-field velocity and density fields due to the simulated free-swimming copepods are calculated from the unsteady Navier-Stokes equations in the Boussinesq limit for an incompressible Newtonian fluid coupled with an advection-diffusion equation. Specific details of our methodology can be found in \citet{ouillon2020active}. In brief, the dimensional form of the governing equations is given by
\begin{equation}
  \frac{\partial \boldsymbol u}{\partial t}+\nabla\boldsymbol\cdot(\boldsymbol u\boldsymbol u)=-\frac{1}{\rho_0}\nabla p+\nu_f\nabla^2\boldsymbol u+\boldsymbol f_{IBM}+\xi_f(\boldsymbol x)c\alpha\boldsymbol g,
  \label{NSE}
\end{equation}
\begin{equation}
  \nabla\boldsymbol\cdot\boldsymbol u=0,
  \label{CE}
\end{equation}
\begin{equation}
  \frac{\partial c}{\partial t}+\nabla\boldsymbol\cdot(\boldsymbol {\hat{u}}c)=\kappa_f \nabla^2 c,
  \label{ad}
\end{equation}
where $\boldsymbol u=(u,v,w)^T$ designates the fluid velocity vector in Cartesian components and $\boldsymbol{\hat{u}}$ is the compound velocity. It is defined as the volume-weighted average of the velocity of the particle, which is the solid body motion of the rotating and translating swimmer, and the fluid in the vicinity of the particle itself. This ensures no advective concentration transport inside the spherical particle \citep{ouillon2020active}. Properties of the fluid are given by $\rho_0$, the reference density, $\nu_f$, the kinematic viscosity, and $p$, the pressure. Other parameters include the time $t$, the gravitational acceleration $\boldsymbol g$, and the artificial volumetric force introduced by the Immersed Boundary Method (IBM), $\boldsymbol f_{IBM}$  \citep{uhlmann2005immersed,biegert2017collision}. The volumetric force acts on the surface of the particle, effectively linking the motion between the particle and the fluid phases. Hence, $\boldsymbol u$ in Eq. 4 is used to calculate this forcing term. In Eq. 5, $\alpha$ is the expansion coefficient associated with the concentration field $c$. In Eq. 7, $\kappa_f$ is the scalar diffusivity. Here, we assume a linear relationship between the local density $\rho_f$ and the scalar concentration,
\begin{equation}
  \rho_f = \rho_0(1+\alpha c).
  \label{den}
\end{equation}
Finally, $\xi_f(\boldsymbol x)$ is the indicator function of the fluid phase,
\begin{eqnarray}
\xi_f(\boldsymbol x)&=&\left\{
\begin{array}{ll}
1 & \ \mathrm{if}\ \boldsymbol x\in \Omega_f,\\
0 & \ \mathrm{if}\ \boldsymbol x\in \Omega_p,
\end{array} \right. \label{svpl}
\end{eqnarray}
where $\Omega_f$ and $\Omega_p$ indicate the volumes occupied by the fluid and swimmer phases, respectively.  We employ the Volume of Fluid (VoF) approach to track the concentration field such that scalars do not overlap with the swimmer phase.

\subsection{Non-dimensional formulation} \label{sec.Ndf}
We non-dimensionalize the governing equations by considering the swimmer radius $R_p$ as the characteristic length scale. The characteristic velocity scale, also defined as the composite velocity, is taken as the sum of the terminal velocity of a squirmer in the Stokes regime \citep{blake1971spherical} and the Stokes settling velocity of a swimmer at the reference density of the fluid \citep{doostmohammadi2014numerical}, $U_{comp}=\frac{2}{3}B_1+\frac{2}{9}\frac{(\rho_p-\rho_0)gR^2_p}{\rho_0\nu_f}$. The characteristic concentration $C$ is chosen to make the dimensionless concentration vary between 0 and 1. The non-dimensional variables are given by
\begin{eqnarray}
\nonumber  &\boldsymbol x=R_p\tilde{\boldsymbol x},\ \boldsymbol u=U_{comp}\tilde{\boldsymbol u},\ t=\frac{R_p}{U_{comp}}\tilde{t},\ p=\rho_0 U^2_{comp}\tilde{p},\\
  &\boldsymbol f_{IBM}=\rho_0 U^2_{comp}R_p^2\tilde{\boldsymbol f}_{IBM},\ m_p=\rho_0 R^3_P\tilde{m}_p,\ \ V_p=R^3_p\tilde{V}_p,\\ 
\nonumber &\ I_p=\rho_0 R^5_p\tilde{I}_p, \ \boldsymbol \omega_p=\frac{U_{comp}}{R_p}\tilde{\boldsymbol \omega}_p,\ c=C\tilde{c},
\label{NDCS} 
\end{eqnarray}
where dimensionless quantities are denoted by a tilde. Here, $\boldsymbol x$ represents any length and $\boldsymbol u$ is any velocity vector. In this way, we obtain the dimensionless equations,
\begin{eqnarray}
&\frac{\partial \tilde{\boldsymbol u}}{\partial \tilde{t}}+\tilde{\nabla}\boldsymbol\cdot(\tilde{\boldsymbol u}\tilde{\boldsymbol u})=-\tilde{\nabla}\tilde{p}+\frac{1}{Re}\tilde{\nabla}^2\tilde{\boldsymbol u}+\tilde{\boldsymbol f}_{IBM}+Ri\xi_f \boldsymbol e_g \tilde{c}, \label{NDEm} \\
&\tilde{\nabla}\boldsymbol\cdot\tilde{\boldsymbol u}=0, \label{NDEc-1} \\
&\frac{\partial \tilde{c}}{\partial \tilde{t}}+\tilde{\nabla}\boldsymbol\cdot(\tilde{\boldsymbol u}\tilde{c})=\frac{1}{Pe}\tilde{\nabla}^2\tilde{c}, \label{NDEc-2} \\
&\tilde{m}_p\frac{\mathrm{d}\tilde{\boldsymbol u}_p}{\mathrm{d}\tilde{t}}=\oint_{\tilde{\Gamma}_p}\tilde{\boldsymbol\tau}\boldsymbol\cdot\boldsymbol n\ \mathrm{d}\tilde{A}+\tilde{V}_p\frac{\rho_p-\rho_f}{\rho_0}G\boldsymbol e_g,\\
&\tilde{I}_p\frac{\mathrm{d}\tilde{\boldsymbol \omega}_p}{\mathrm{d}\tilde{t}}=\oint_{\tilde{\Gamma}_p}\tilde{\boldsymbol r}\times(\tilde{\boldsymbol\tau}\boldsymbol\cdot\boldsymbol n)\ \mathrm{d}\tilde{A},\label{NDEa}
\end{eqnarray}
where $\boldsymbol{e}_g$ is the unit vector in the direction of gravity. The governing dimensionless parameters are $Re=U_{comp}R_p/\nu_f$, $Ri=gC\alpha R_p/U^2_{comp}$, $Pe=U_{comp}R_p/\kappa_f$, and $G=g R_p/U^2_{comp}$ The non-dimensional squirmer velocity boundary condition becomes $u_{\theta}=\frac{B_1}{U_{comp}}(\mathrm{sin}\theta+\beta\mathrm{sin}\theta\mathrm{cos}\theta)$. For convenience, the tilde symbol will be omitted henceforth.

\subsection{Numerical set-up}
Simulations are conducted with our \emph{particle-laden flows via immersed boundaries} (PARTIES) in-house code. \citep{biegert2017collision}. We employ a third-order low-storage Runge-Kutta (RK) scheme and a second-order accurate finite difference approach in time and space, respectively \citep{biegert2017collision}. The pressure-projection method is implemented based on a direct fast Fourier transform solver for the resulting Poisson equation at each RK sub-step. The fluid velocity and concentration field boundary conditions at the surface of the swimmers are enforced via the IBM and VoF techniques.

The values of the governing parameters, such as body length, body orientation, swimming speed, and Reynolds number are closely matched to lab measurements. Based on experimental observations, we select the swimmer radius $R_p\sim 0.25$ mm, and the swimming speed $u_p \sim 1\  \mathrm{ mm\ s^{-1}}$. For a fluid viscosity $\nu_f$ of $10^{-6}\ \mathrm{m^2\ s^{-1}}$, this yields a Reynolds number $Re \sim$ 0.25. It should be noted that the definition of $Re$ here differs slightly from that employed in experiments. While experimental studies generally adopt the body length as the characteristic length scale, the present numerical simulations utilize the sphere radius for computational convenience. Consequently, the Reynolds numbers reported herein are approximately half of the values reported in the corresponding experimental benchmarks. The body density of the swimmer is around 1003 kg m$^{-3}$. Varying the direction of the initial relative squirming velocity $\boldsymbol e_{\theta}$ for different cases, we can obtain different body orientations.  

\section{Results}

\subsection{Experimental observations}\label{Experimetnal results}
\subsubsection{Copepod swimming and free-sinking speeds}
\label{individual copepod swimming speed}

\begin{figure}
    \centering
    \includegraphics[width = 1\textwidth]{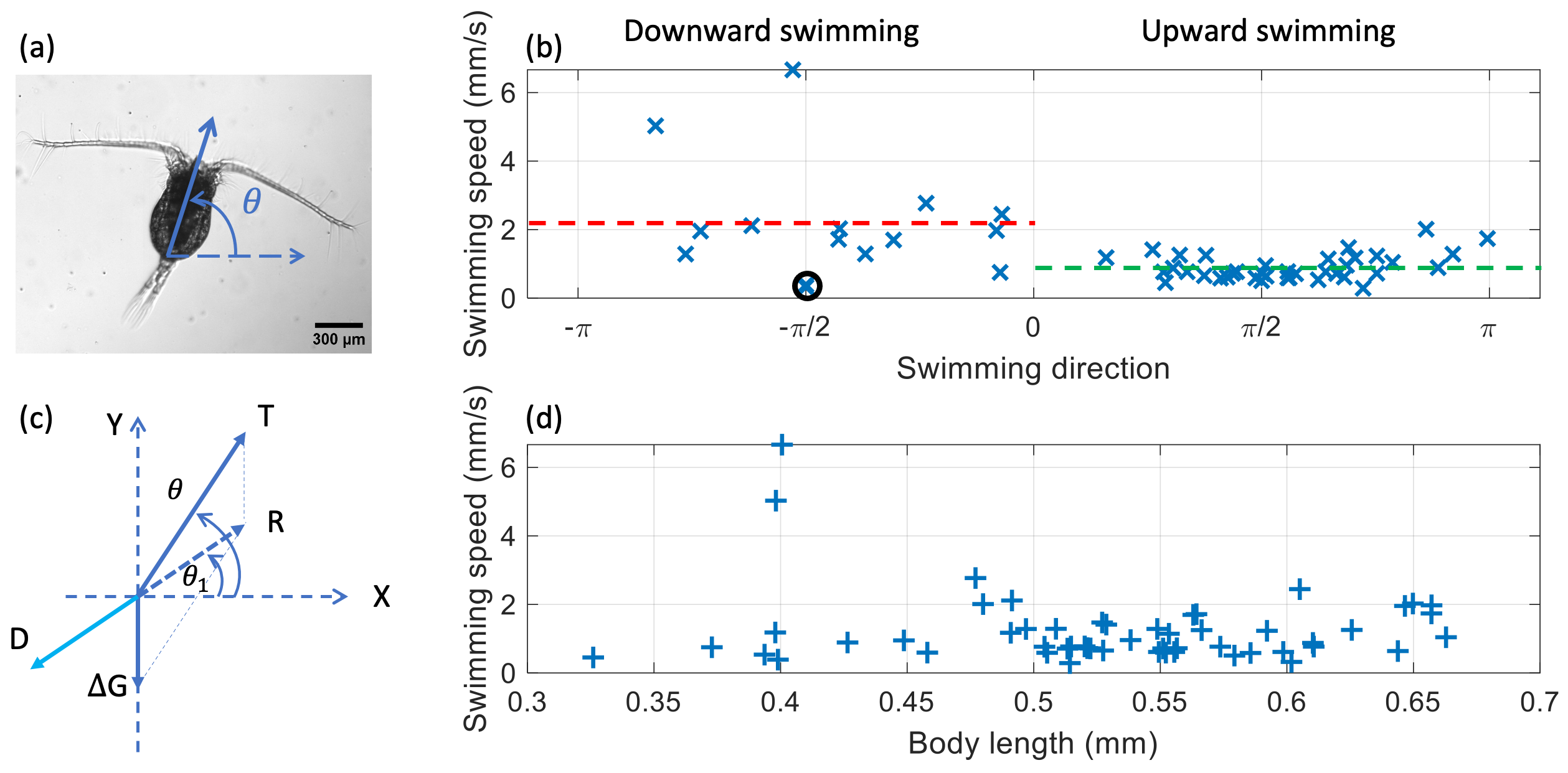}
    \caption{Swimming speed measurements of individual copepods. Panel (a) shows the copepod swimming angle, $\theta$, with respect to the horizontal direction, with negative values for downward swimming and positive values for upward swimming. Note that the focal plane is vertical. Panel (b) shows the swimming speed at different swimming angles as defined in panel (a). The dashed lines indicate the averaged upward (green) and downward (red) swimming speed, respectively. The black circle denotes the freely sinking speed of a copepod.  Panel (c) shows the free-body diagram of a swimming copepod at steady state, where $T$ is the thrust force, $\Delta G$ is the excess weight, $D$ is the drag force, $R$ is the resultant force of gravity and thrust. The angle between the resultant force, $R$, and the horizontal direction is denoted as $\theta_1$. Panel (d) shows the measured individual copepod swimming speeds as a function of organism body length.   }
    \label{fig:indiv copepod speed}
\end{figure}

We conducted experiments to measure the swimming speed of different individual copepods (n = 54 measurements), defined as the average organism cruising speed during forward motion. This metric is a function of swimming direction and copepod body length (Fig.~\ref{fig:indiv copepod speed}). Here, the swimming angle is defined using the horizontal plane (dashed line in Fig.~\ref{fig:indiv copepod speed}(a)) and the organism body axis, in which the upward direction is considered positive.

The average downward swimming speed was observed to be considerably higher than the average upward speed (red and green dashed lines, respectively, in Fig.~\ref{fig:indiv copepod speed}(b)), with the individual copepod swimming speed being independent of organism body length (Fig.~\ref{fig:indiv copepod speed}(d)). A similar trend was observed across a broad range of field measurements of mesozooplankton aggregations performing DVMs \citep{klevjer2012distribution,bianchi2016global,tarling2001swarm,ashjian2002distribution}. In particular, \citet{bianchi2016global} analyzed acoustic Doppler current profiler (ADCP) data from about 3,400 DVM events across nearly 390 cruises between 1990 and 2010. They observed that mesozooplankton migrated faster during their descent to deeper less oxygenated regions (7.6 $\pm$ 3.6 cm s$^{-1}$) than while ascending to the ocean surface (6.5 $\pm$ 3.5 cm s$^{-1}$). This asymmetry in migration speed was attributed to behavioral traits related to anoxic water conditions, physical effects arising from the gravity (excess weight) of the animals \citep{davison2011specific}, and the preference for darker light levels at dawn compared to dusk \citep{staby2011follow}. Here, in the absence of behavioral stressors, the difference in swimming speeds is attributed to the negative buoyancy of the copepods (an effect also recently observed by \citet{tack2024ups}). For negatively buoyant organisms, upward swimming requires additional effort to overcome gravity, while this imbalance can be leveraged when swimming downward. This asymmetry is important as it can lead to different near-body flow fields as well as to different net amounts of BHT, as explored in the following sections.

Besides measuring the copepod swimming speeds attained with different body orientations, we also quantified the mean copepod free-sinking speed (black marker in Fig.~\ref{fig:indiv copepod speed}(b)). From Stokes flow theory and considering the force balance between gravitational and drag forces, the terminal speed of a freely sinking sphere can be expressed as 
\begin{equation}
    v_t = \frac{2}{9}gR_p^2\Delta \rho/\mu, \label{terminal sphere speed}
\end{equation}
where $g$ is the defined standard value for gravitational acceleration, $R_p$ is the radius of the sphere, $\Delta \rho$ is the density difference (net density) between the sphere and the ambient fluid, and $\mu$ is the fluid viscosity. A density difference of $\Delta \rho\sim$3 kg/m$^3$ was computed using this equation by using the measured terminal velocity of the copepods. This parameter was crucial for setting up the squirmer model simulations of individual swimming copepods. 

It is important to note that most of the observed copepods swam at an oblique angle relative to gravity, rather than strictly in the vertical direction ($\theta = -\pi/2$ or $\pi/2$, aligned with gravity,  Fig.~\ref{fig:indiv copepod speed}(b)). The misalignment between the organism body axis and gravity caused the net swimming direction to deviate slightly from the organism orientation. This can be demonstrated with a free-body diagram of a steadily swimming copepod influenced by excess weight, thrust, and drag forces (Fig.~\ref{fig:indiv copepod speed}(c)). The excess weight is due to the density difference between the swimmer and the ambient fluid. The thrust force, generated by the beating appendages, is assumed to align with the body axis. The drag force acts in the direction opposite to the copepod swimming direction, impeding its motion. During steady-state swimming, these three forces achieve equilibrium, resulting in a trigonometric relationship between the excess weight force, $\Delta G$, and the thrust force, $T$,
 \begin{equation}
     T/\Delta G = |\cos(\theta_1)|/\sin(\theta-\theta_1),
     \label{T/G} 
 \end{equation}
where $\theta$ is the angle between the horizontal direction (x-axis) and the body axis orientation (thrust direction), and $\theta_1$ is the angle between the actual swimming direction and the horizontal direction (x-axis). Using the copepod swimming data, we can then estimate the magnitude of the thrust force produced by swimming copepods in different swimming directions. This force analysis will be analyzed in Section~\ref{Sect: force}.

\subsubsection{Near-field body flows} \label{sect: copepod flow fields}

\begin{figure}
    \centering
    \includegraphics[width = 1\textwidth]{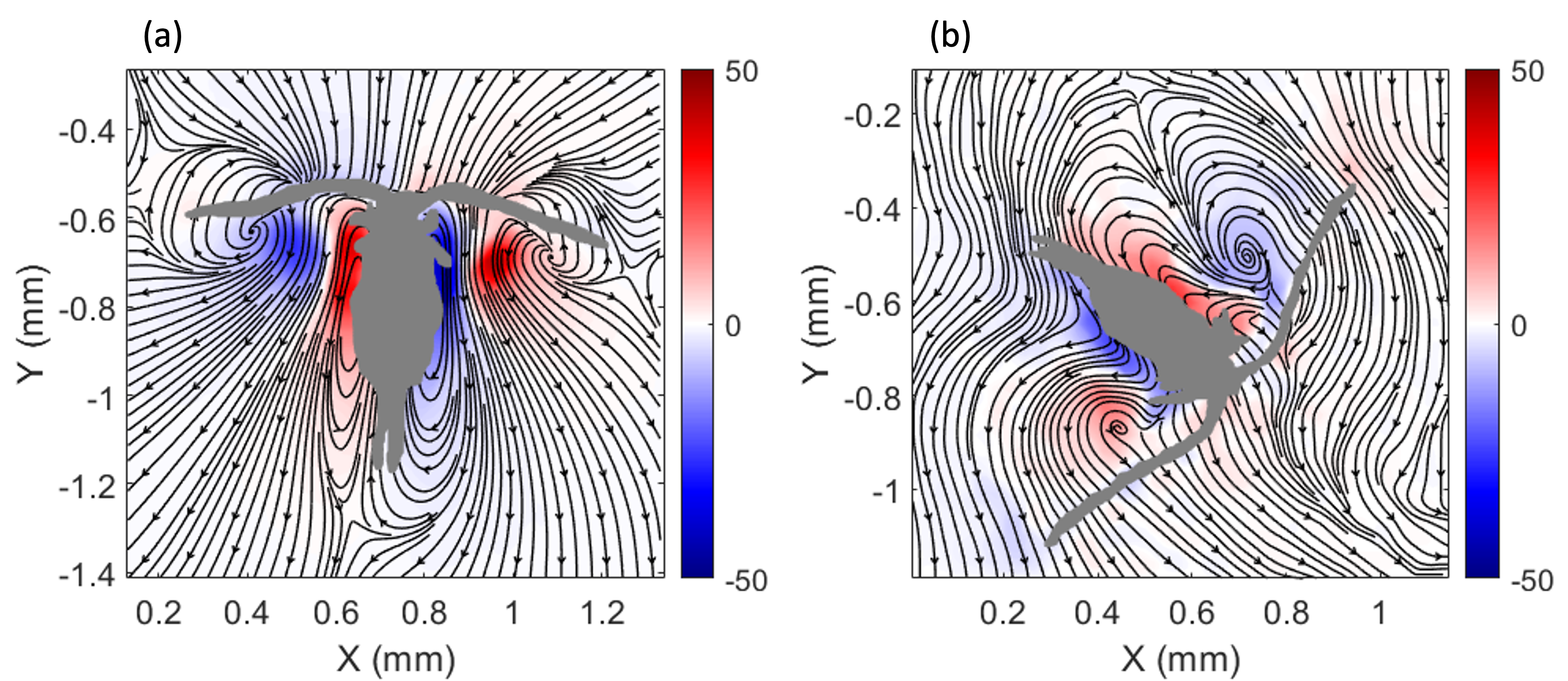}
    \caption{Cycle-averaged vorticity fields ($s^{-1}$) and streamlines of the flow fields induced by individual swimming copepods in the lab frame of reference (gravity direction is in the negative y-direction). Panel (a) shows a front-view (dorso), upward swimming copepod; panel (b) shows a front-view (dorso), downward swimming copepod. } 
    \label{fig:indiv copepod flow fields-top}
\end{figure}

We used bright-field micro PIV to capture the near-body flow features induced by individual copepods. The cycle-averaged flow fields are representative of the flow characteristics generated by individual swimming copepods given the variability introduced by the metachronal  beating of the appendages (Fig.~\ref{fig:indiv copepod flow fields-top}).

Capturing upward-swimming organisms from the front view (dorso) revealed how fluid near the top of the copepod is pulled downward toward the copepod and then pushed away, generating an upward thrust force that is essential for swimming  (Fig.~\ref{fig:indiv copepod flow fields-top}(a)). Recirculation regions are observed on either side of the body near the beating appendages. Similar flow features have been reported in the literature \citep{malkiel2003three,jiang2004hydrodynamics,catton2007quantitative,kiorboe2014flow, tack2024ups}. Recirculating vortex structures are also observed at the rear part of the body, which is attributed to the local flow induced by the upward-moving body despite the general downward appendage-induced flow field. Conversely, in the downward swimming case, the flow field near the copepod body displays a markedly different behavior (Fig.~\ref{fig:indiv copepod flow fields-top}(b)). Instead of generating a backward flow field, as in the upward swimming case, the flow induced by the downward swimming copepod is predominantly aligned with the swimming direction. This change was also observed by \citet{tack2024ups} and indicates that the copepod pushes fluid forward during downward swimming instead of pulling it backward. Similar flow fields were also observed in earlier experimental measurements of different copepod species ({\it Metridia longa} and {\it Podon intermedius}) \citep{kiorboe2014flow}, but were not discussed in detail.

\begin{figure}
    \centering
    \includegraphics[width = 1\textwidth]{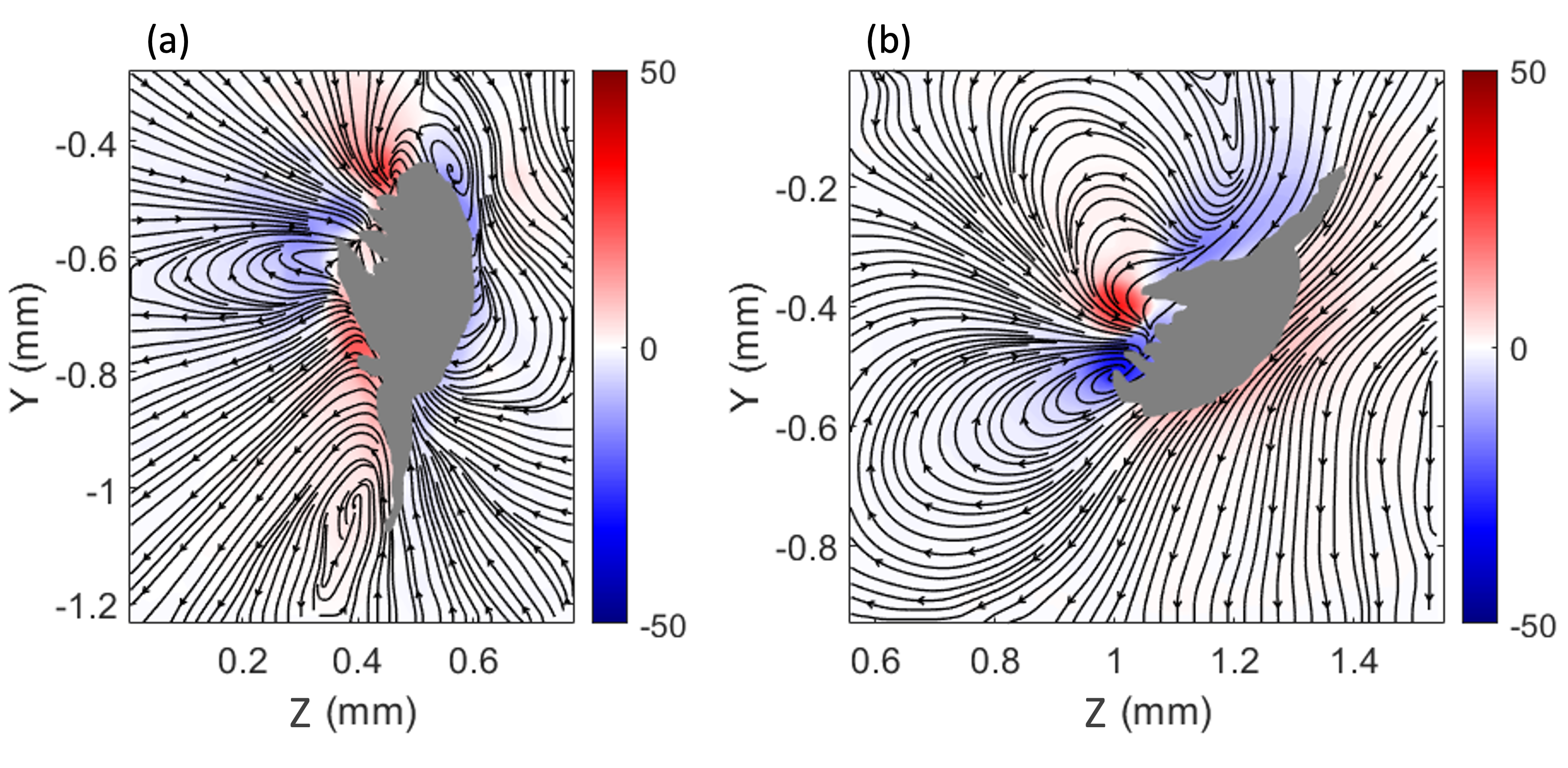}
    \caption{Cycle-averaged vorticity fields (contours, unit $s^{-1}$) and streamlines of the flow fields induced by individual swimming copepods in the lab frame (gravity direction is in negative y direction). Panel (a) shows a side view, upward swimming copepod; panel (b) shows a side view, downward swimming copepod.}
    \label{fig:indiv copepod flow fields-side}
\end{figure}

We also conducted experiments capturing upward and downward swimming organisms from the side  (Fig.~\ref{fig:indiv copepod flow fields-side}). In both cases, the beating appendages induce a strong flow driving fluid towards the body. However, the push-back flow by the beating appendages is significantly stronger in the upward swimming case (Fig.~\ref{fig:indiv copepod flow fields-side}a) compared to the downward swimming case (Fig.~\ref{fig:indiv copepod flow fields-side}(b)). Typically, the beating appendages of a swimming organism produce a backward flow to obtain thrust for locomotion, which is in line with the flow field generated by an upward swimming copepod (Fig.~\ref{fig:indiv copepod flow fields-side}(a)). However, in the downward swimming case (Fig.~\ref{fig:indiv copepod flow fields-side}(b)), the flow velocity in front of the body is aligned with its swimming direction. This forward flow around the copepod body is consistent with the observations in the front-view flow fields (Fig.~\ref{fig:indiv copepod flow fields-top}(b)). From the side-view flow fields (Fig.~\ref{fig:indiv copepod flow fields-side}), we can clearly see the interplay between the appendage-induced flow and the body-drifting flow (the flow caused by body motion), especially in the downward swimming case (Fig.~\ref{fig:indiv copepod flow fields-side}(b)).

Based on the front- and side-view flow fields presented in this section, two fluid transport mechanisms emerge during forward motion. The first mechanism originates from the beating appendages, which drive the flow opposite to the swimming direction. The second mechanism is related to the body of the organism, causing the fluid to drift along with it as the copepod swims. These two competing effects ultimately shape the observed flow field and govern the net transport of fluid in the near-field of the organisms. Notably, our assessment agrees with the theoretical results presented in \citet{jiang2002flow_I}, in which the beating appendages are modeled as a force dipole and the organism body is modeled as a sphere in the Stokes flow regime. Building on the insights of this study, we aim to analyze the dynamics of downward swimming--a distinct mode of locomotion yet to be analyzed in detail.

By comparing between the near field flows during ascent and descent, it is evident that, when swimming upward, copepods need to energetically beat their appendages to generate thrust and counteract both the excess weight and the drag force. As a result, the flow induced by their appendages prevails over the body-drifting flow, leading to a general downward flow near the organism. In contrast, when swimming downward, negative buoyancy (due to their excess weight) is leveraged to achieve greater swimming speeds (Fig.~\ref{fig:indiv copepod speed}(b)) and more substantial body-drifting flow. Consequently, the flow induced by the beating appendages no longer dominates. These combined effects produce a flow field that aligns with the downward swimming direction (Figs.~\ref{fig:indiv copepod flow fields-top}(b),\ref{fig:indiv copepod flow fields-side}(b)). The recent experimental study by \citet{tack2024ups} reported similar flow fields by downward swimming copepods ({\it T. longicornis}). However, the organisms in their study displayed a quantifiable difference in the copepod swimming kinematics--an effect not observed here using a different copepod species. Important to note is that while gravitational effects (excess weight or buoyancy) act in the vertical direction, the drag force is not necessarily following that orientation, leading to either enhancing or canceling effects depending on the swimming direction. This will be further discussed in Section~\ref{sect: drift volume}.

\begin{figure}
    \centering
    \includegraphics[width = 1\textwidth]{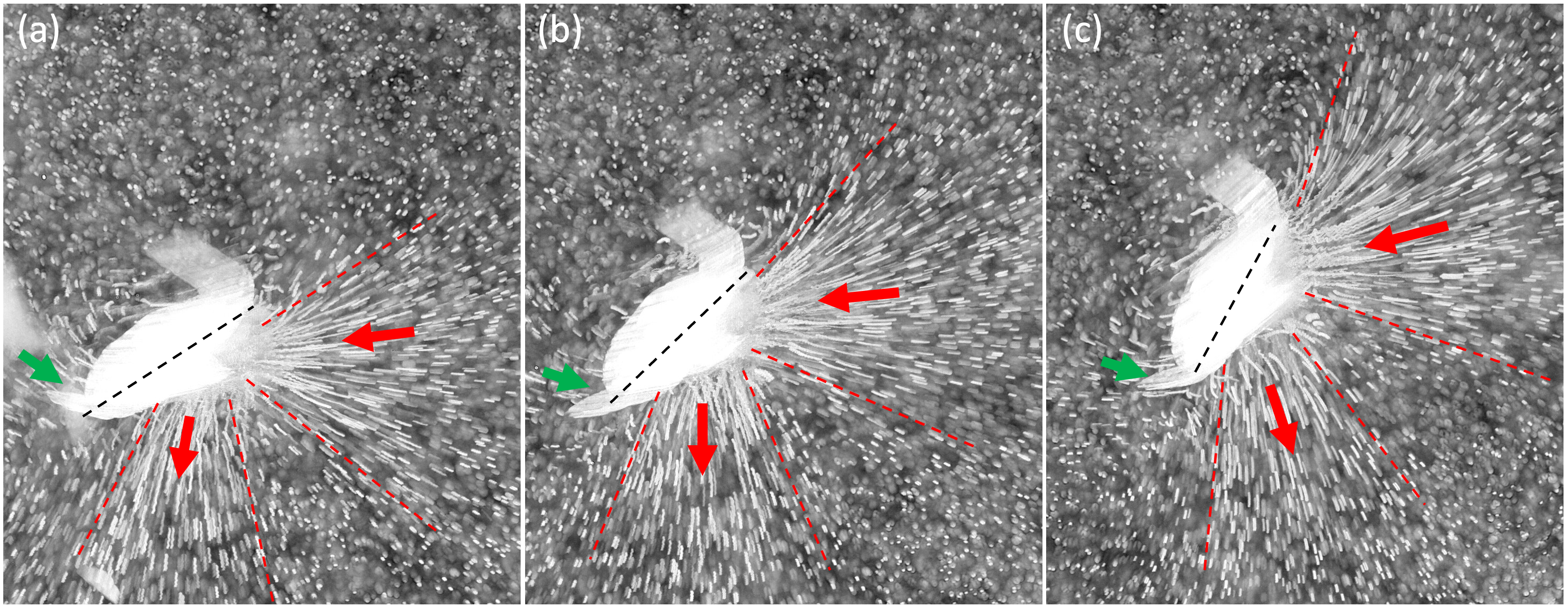}
    \caption{Near-field flow. The time-averaged image sequence in panels (a-c) shows the particle streaks induced by a swimming copepod (side view, with gravity pointing downward). In this sequence, from (a) to (c), the copepod is changing its swimming direction from an approximately horizontal direction to an approximately vertical direction. The red and green arrows denote the general direction of the seed particles and the near-body drift flow direction, respectively. The black dashed lines denote the body axis of the copepod, and the red dashed lines show the range of the flow induced by the organism.}
    \label{fig:up vs down particles}
\end{figure}

The aforementioned interplay between appendage-driven and body-drifting flows becomes evident as a forward swimming copepod transitions from horizontal to upward motion (Fig.~\ref{fig:up vs down particles}). As an organism changes orientation, the copepod induces a strong convergent inflow of tracer particles near its anterior at a slight angle while a strong divergent outflow is concurrently generated near the rear part of the organism. Throughout the change in orientation, the body-drifting flow (green arrows in Fig.~\ref{fig:up vs down particles}) remains relatively small compared to the appendage-induced flow (red arrows in Fig.~\ref{fig:up vs down particles}). Similar convergent influx and divergent outflux flow patterns have been reported in the literature of both numerical studies of simplified swimmers \citep{jiang2002flow_II} and experimental studies of copepods \citep{strickler1982calanoid,malkiel2003three,catton2007quantitative}. It is worth highlighting that the relative size of the influx flow increases as the copepod shifts its swimming direction from horizontal to vertical, suggesting that the swimmer is generating a stronger upward thrust force to overcome the excess weight effects when swimming upward. The flow fields in panels (a-c) of Fig.~\ref{fig:up vs down particles} demonstrate how excess weight results in large downward flows by upward swimming organisms.

\subsection{Discrete swimmer simulations}\label{simulation results}

\subsubsection{Validation of numerical setup}

\begin{figure}
    \centering
    \includegraphics[width = 1\textwidth]{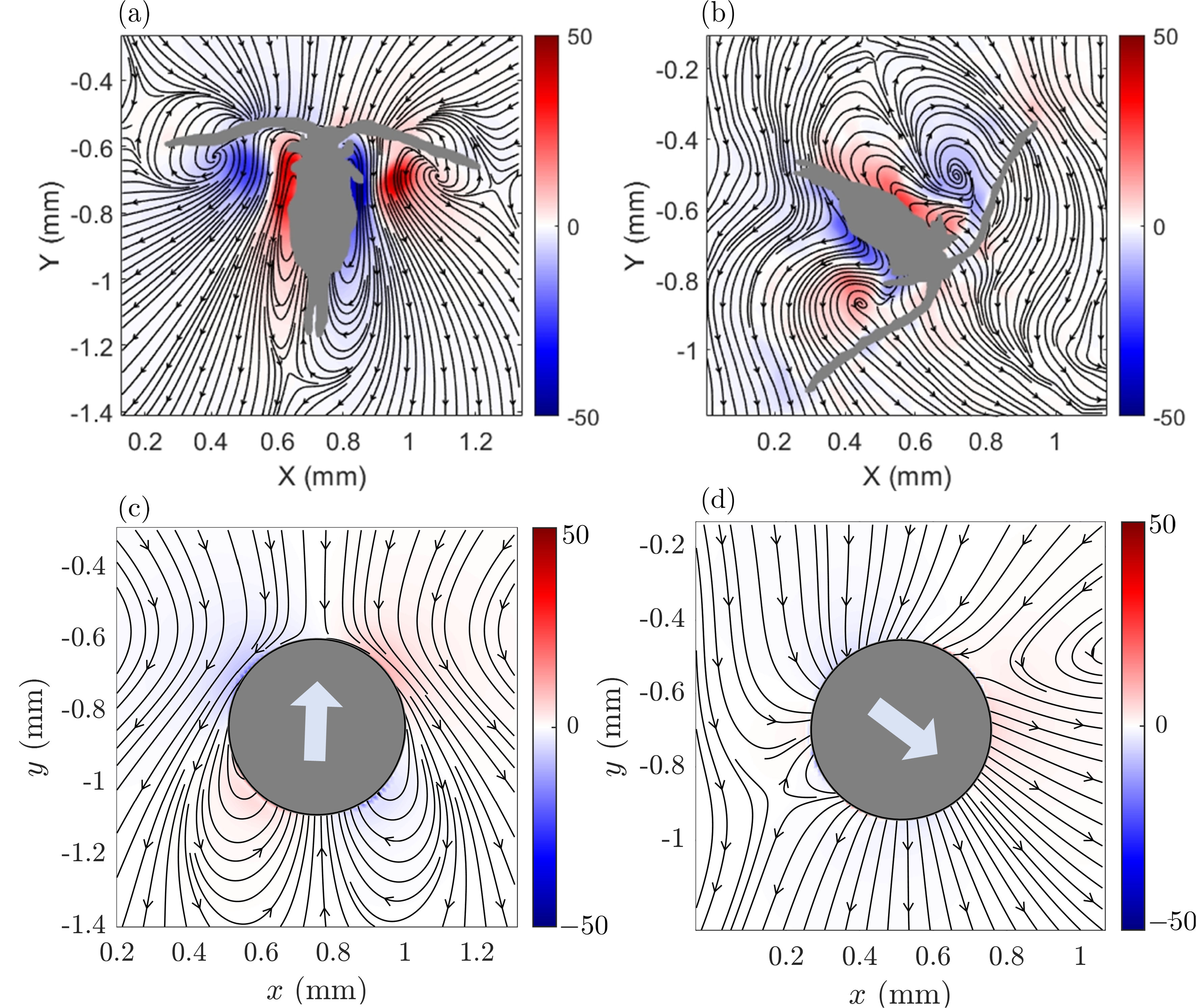}
    \caption{Induced flow fields due to simulated and real free-swimming organisms. The contours represent vorticity fields (in s$^{-1}$), while the streamlines with arrows depict the flow field. Panels (a) and (c) present a front view of an upward-swimming copepod (case V1). Panels (b) and (d) display a front view of a downward-swimming copepod (case V3). The arrows in panels (c) and (d) indicate the swimming direction.}
        \label{fig: Validation_ff}
\end{figure}

We validate our numerical approach by comparing simulation results with the corresponding experimental data for both the upward and the downward swimming cases. From the lab experiments, we quantify the organism body length, the swimming speed, the body orientation, and the swimming direction. In the simulation setup, we fix the body length and body orientation to match three of the copepods in the experiments. Each run is then repeated several times with different values of the first propulsion mode $B_1$, the squirming mode $\beta$, and the particle density $\rho_p$ until we observe good quantitative agreement with the experimental values of organism mean swimming speed and swimming direction. It should be noted that the exact swimmer density is not available. As outlined in Section \ref{sec:Mo}, the copepod species is reported to be slightly negatively buoyant. The initial configuration, along with the simulated swimming speed and direction are listed in Table \ref{Tab.Ces}. In addition, we also compared the lab and simulated flow fields in the vicinity of the free-swimming organisms (Fig. \ref{fig: Validation_ff}). Given that the individual copepods are being modeled as spherical particles, the near-field is not discernible in the simulation results. Nonetheless, the simulated flow field at a slight distance from the body is in good agreement with the experimental observations. Details on the validation results for PARTIES are presented in earlier work by \citet{biegert2017collision,ouillon2020active} and \citet{zhu2022grain}.

\begin{table}
  \begin{center}
\def~{\hphantom{0}}
  \begin{tabular}{lccc}
       Run &  V1 & V2 & V3     \\[3pt]
       Particle diameter $d_p\ \mathrm{(mm)}$ & 0.56  & 0.43  &  0.46 \\
       Particle orientation $\theta$ (rad)  & 1.53     & 2.39     &  -0.65 \\
       Particle density $\rho_p\ \mathrm{(kg\ m^{-3})}$  & 1003  & 1002  & 1003 \\
       First mode (propulsion) $B_1\ \mathrm{(mm\ s^{-1})}$ & 1.5 & 0.9    & 1.4 \\
       Squirming mode  $\beta$ & 1 & 1  & -0.5 \\
       Experimental mean swimming speed (mm s$^{-1}$) & 0.68 & 0.42   &  1 \\
       Simulated mean swimming speed (mm s$^{-1}$) & 0.63 & 0.48   &  1 \\
       Experimental swimming direction (rad) & 1.53 & 2.67  & -0.83 \\
       Simulated swimming direction (rad) & 1.57  & 2.61  &  -0.84   \\
  \end{tabular}
  \caption{Comparison of the experimental and simulated cases and associated parameter values for three representative experimental test organisms.}
  \label{Tab.Ces}
  \end{center}
\end{table}

\subsubsection{Drift volume analysis} \label{sect: drift volume}
\begin{figure}
    \centering
    \includegraphics[width = 1\textwidth]{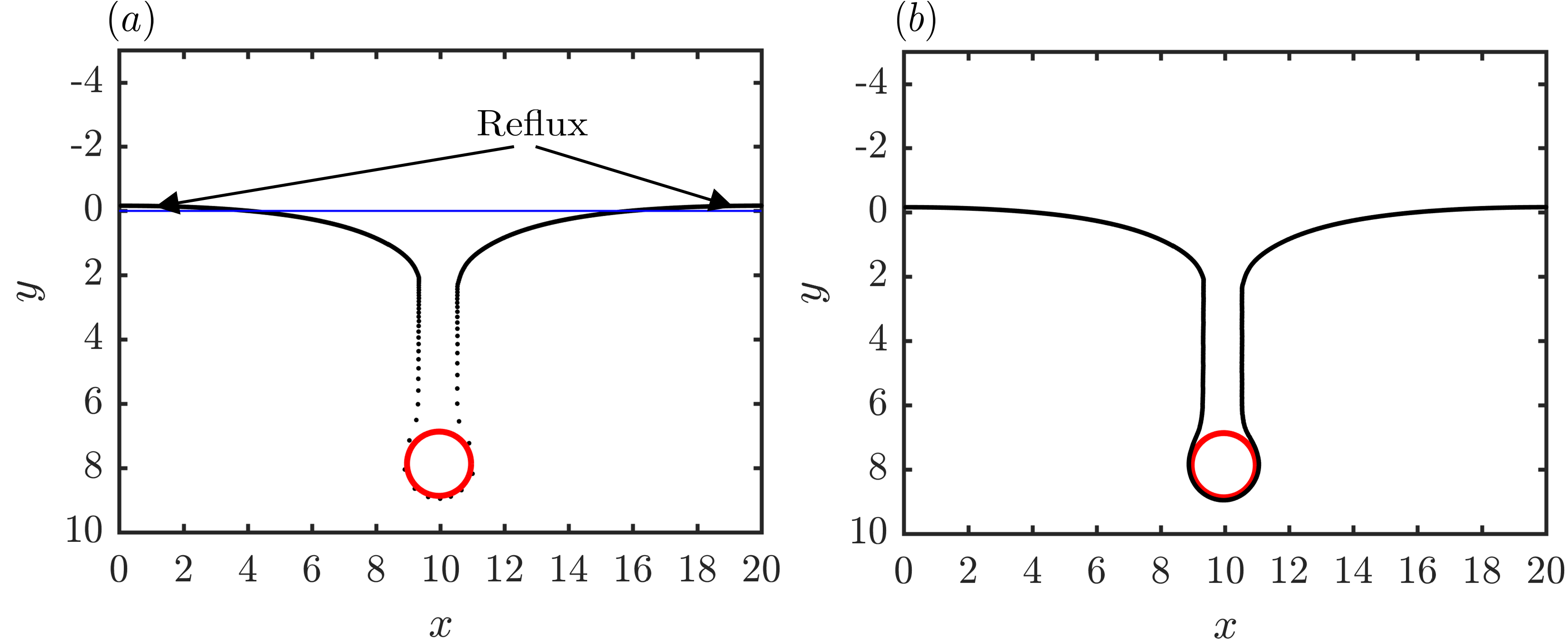}
    \caption{Drift by a spherical swimmer tuned with lab-based observations of free-swimming copepods. Panel (a) shows the positions of Lagrangian tracer particles as the swimmer translates vertically downward a distance of 23 unit lengths. Panel (b) shows the re-meshed locations of the Lagrangian tracer particles shown in panel (a). The red circle shows the position of the spherical swimmer. The black dots are the Lagrangian tracer particles. The blue line indicates the initial position of the Lagrangian tracer plane.}
        \label{fig: drift volume set}
\end{figure}

Drift volume refers to the amount of fluid permanently displaced by the motion of an object in a fluid and has been a key metric for understanding the role of swimming organisms in transporting nutrients, carbon, and oxygen in marine ecosystems \citep{darwin1953note, doostmohammadi2014numerical,katija2015morphology, wilhelmus2019effect}. In this study, we analyze the drift patterns due to simulated swimmers oriented along the vertical direction. Following studies in the literature \citep{young2010effect,doostmohammadi2014numerical}, a horizontal row of Lagrangian tracers is placed 15 units above the swimmer along the vertical ($y$) axis at the beginning of the simulation. Here, one unit corresponds to a dimensionless length of unity. The tracer particles are evenly spaced every 0.02 units along the horizontal ($x$) axis and aligned with the center of the swimmer in the $z$-direction. The motion of the tracers is tracked by solving the initial value problem $\mathrm{d}\boldsymbol x/\mathrm{d}\boldsymbol t=\boldsymbol u$ with $\boldsymbol x(t_0)=\boldsymbol x_0$, in which $\boldsymbol u$ is the fluid velocity at the given position of a tracer particle. We use the Eulerian solution to calculate the position of the tracer particle with the form $\boldsymbol x_{n+1}=\boldsymbol x_n+\boldsymbol u\delta t$, where $\delta t$ is the time step in the simulation. As the swimmer moves forward, the spatial distribution of the Lagrangian tracer particles delineates the surface of the drift volume (Fig. \ref{fig: drift volume set}(a)). We also reconstructed the continuous drift surface by re-meshing the Lagrangian tracer plane (Fig. \ref{fig: drift volume set}(b)). In this case, we supply one particle in between two adjacent tracer particles at each time step if their distance is larger than 0.02 unit lengths. For the remainder of this section, we do not employ the re-meshed drift surface to focus on the drift volume by the swimmer as it travels in a finite domain. As the swimmer moves 30 unit lengths vertically, the volume between the distorted Lagrangian tracer plane and its original position is defined as the drift volume $V_D$, including the magnitude of the reflux volume (Fig. \ref{fig: drift volume set}(a)). Given its symmetry about the $y$-axis, the drift volume $V_D$ is computed by revolving the swimmer outline and drift surface about the $y$-axis by $\pi$ rad.

\begin{figure}
    \centering
    \includegraphics[width = 1\textwidth]{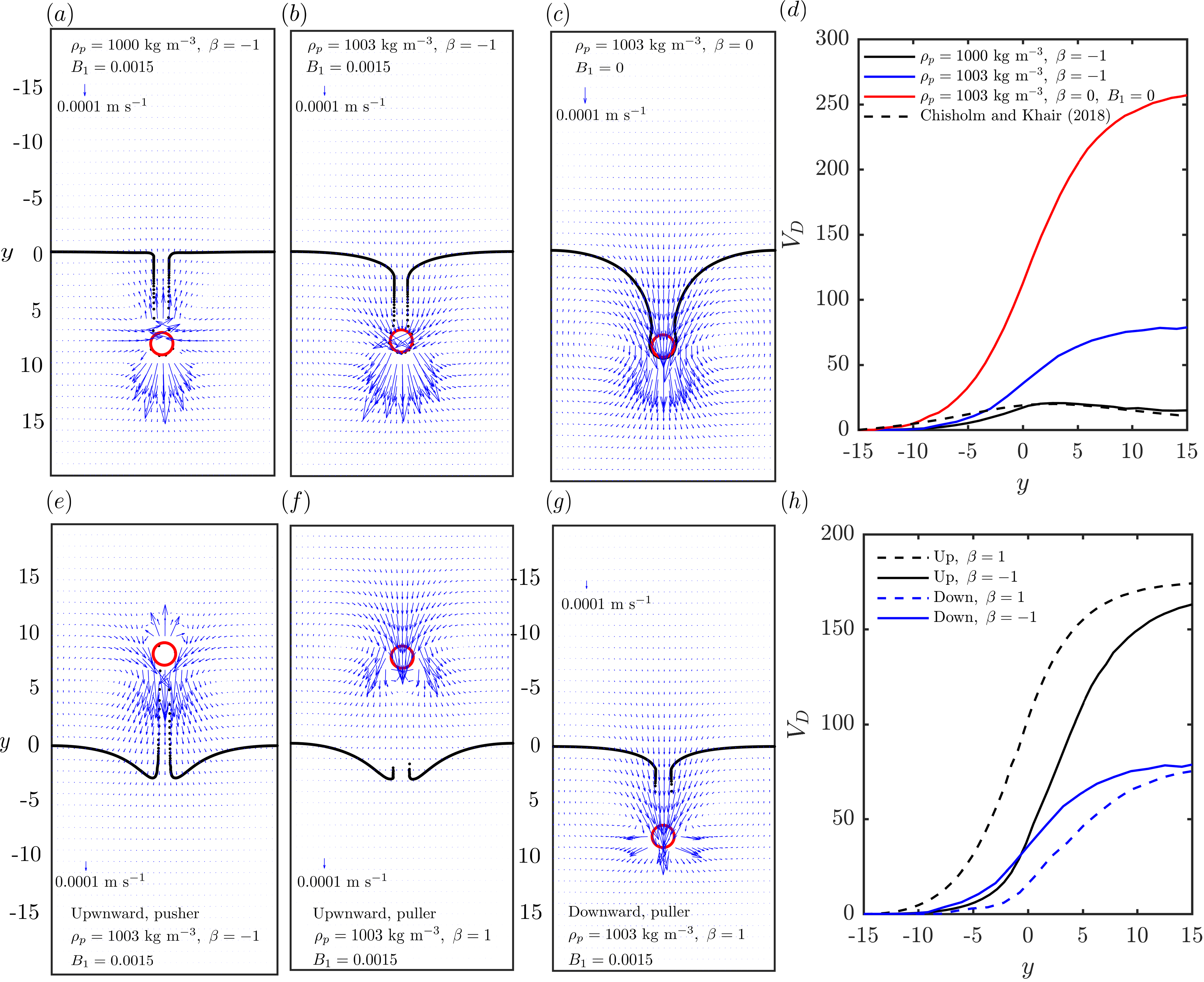}
    \caption{Drift by vertical swimmers in homogeneous fluid. Panels (a-c) and (e-g) show the shape of the drift surface in black and the flow field around the swimmer with blue arrows for both neutrally buoyant swimmers (a) and negatively buoyant cases (rest of panels). Top panels represent the neutrally buoyant swimmer, negatively buoyant swimmer, and passively settling particle; while bottom panels represent upward and down swimmers with different modes (pushers and pullers). Panels (d) and (h) show the spatial evolution of the net drift volume $V_D$ as a function of the position of the swimmers. The black dashed line in (d) and (h) reproduces the theoretical results of a neutrally buoyant swimmer presented by \citet{chisholm2018partial}.}
        \label{fig: drift volume}
\end{figure}

Downward swimming cases were parameterized to match experimental estimates, in which $d_p=0.56$ mm, $B_1=1.5\ \mathrm{mm\ s^{-1}}$,  and $\beta=-1$, yielding $Re=0.28$. Alongside a realistic swimmer density ($\rho_p=1003\ \mathrm{kg\ m^{-3}}$), we tested a neutrally buoyant swimmer ($\rho_p=1000\ \mathrm{kg\ m^{-3}}$) and a passively settling particle with the same density as the realistic swimmer density for which $B_1=0$ and $\beta=0$. For the neutrally buoyant swimmer, tracer particles initially ahead of the body are displaced downward by the beating appendages and then pulled backward by an upward flow behind the body (Fig. \ref{fig: drift volume}(a)). Consequently, the drift volume peaks as the swimmer crosses the interface at $y \sim 0$ (black solid line in Fig.~\ref{fig: drift volume}(d)). As discussed in Sect.~\ref{sect: copepod flow fields}, the lack of excess weight effects in this neutrally buoyant case result in a slightly fore-aft asymmetric velocity distribution and a minimal net contribution to the drift volume $V_D$, consistent with previous theoretical findings (e.g., \cite{leshansky2010small,subramanian2010viscosity,pushkin2013fluid,chisholm2018partial}). Note that we characterize asymmetry by the spatial distribution of flow directions, independent of local variations in velocity magnitude.

A passively settling particle, in contrast, induces strong downward flows both upstream and downstream of the body, resulting in a more pronounced fore-aft asymmetry (Fig.~\ref{fig: drift volume}(c)). The drift volume is large given the significant amount of fluid being transported downward (red solid line in Fig.~\ref{fig: drift volume}d). For negatively buoyant swimmers (Fig.~\ref{fig: drift volume}(b)), the net drift volume (blue line in Fig.~\ref{fig: drift volume}(d)) remains bounded between the neutral and passive settling cases. While excess weight and drag both contribute to the fore-aft asymmetry of the flow, these two forces act in opposition leading to a canceling effect that mitigates net fluid displacement. 

Considering all three cases, the effect of any fore-aft asymmetry on net transport becomes evident. The jet generated in the wake of the neutrally buoyant swimmer balances the upstream flow as the swimmer descends (Fig.~\ref{fig: drift volume}(a)). In contrast, the lack of active propulsion in the case of a settling particle leads to a strong downward flow everywhere around its body (Fig.~\ref{fig: drift volume}(c)). The excess weight experienced by negatively buoyant swimmers during descent (Fig.~\ref{fig: drift volume}(b)) seems to increase the fore-aft asymmetry observed in the neutrally buoyant cases, thereby increasing the amount of net drift attained. Finally, the downward flows generated by the simulated swimmers during descent (Figs.~\ref{fig: drift volume}(b)) align with the experimentally resolved flow fields induced by downward swimming copepods (Figs.~\ref{fig:indiv copepod flow fields-top}(b), \ref{fig:indiv copepod flow fields-side}(b)).

We also simulated ascending swimmers with two different modes (pushers and pullers) and quantified the drift volume $V_D$ for these cases (Fig. ~\ref{fig: drift volume} (e) and (f)). Notably, the drift volumes for upward swimming cases (black lines in Fig.~\ref{fig: drift volume}(h)) are significantly larger than those for their downward swimming counterparts (blue lines in Fig.~\ref{fig: drift volume}(h)). This disparity can be attributed to the excess weight and its role in reinforcing a downward flux in the vicinity of the swimmer. During ascent, a negatively buoyant swimmer must compensate its excess weight, which is acting in the same direction of the induced drag force, to maintain steady locomotion. It is important to highlight that the swimming direction is a primary determinant of the net drift volume attained. The insensitivity of the drift volume to swimming mode is particularly evident when comparing downward- against upward-swimming cases (Fig.~\ref{fig: drift volume}(h)). Transport seems to be dominated by the net distribution of forces rather than the specific distribution of active stresses along the body during active propulsion.

\subsubsection{Production of propulsion force}\label{Sect: force}
As noted in Section~\ref{individual copepod swimming speed}, the force distribution of free-swimming copepods is governed by their excess weight $\Delta G$, thrust $T$, and drag $D$. In equilibrium, we can use the trigonometric relationship in Eq. \ref{T/G} to relate $\Delta G$ and $T$. Here, we analyze that force distribution by conducting a series of simulations varying swimmer density $\rho_p$, swimmer size $d_p$, and swimming mode parameters $\beta$ and $B_1$ (Table \ref{Tab.Osc}).   

\begin{table}[h!]
  \begin{center}
\def~{\hphantom{0}}
  \begin{tabular}{lccccc}
       Run & $\rho_p\ \mathrm{(kg\ m^{-3})}$ & $d_p\ \mathrm{(10^{-3}\ m)}$ & $B_1\ \mathrm{(10^{-3}\ m\ s^{-1})}$ & $\beta$ & $T\ \mathrm{(10^{-9}\ N)}$\\[3pt]
       $1-3$   & 1002, 1003, 1004    & 0.434  & 0.9 & 1 & 3.03, 3.06, 3.04    \\
       $4-7$   & 1002   & 0.35, 0.5, 0.6, 0.7 & 0.9 & 1 & 2.44, 3.75, 4.45, 5.22    \\
       $8-10$   & 1005   & 0.434 & 1.1, 1.5, 2.3 & 1.5 & 3.61, 5.32, 8.47    \\
       $11-12$   & 1005  & 0.434 & 1.5 & 0.75, 3 & 5.44, 5.21    \\
       $13-15$   & 1001, 1002, 1003    & 0.46  & 1.35 & -0.75 & 5.54, 5.48, 5.54     \\
       $16-19$   & 1003   & 0.35, 0.55, 0.65, 0.75 & 1.35 & -0.75 & 4.02, 6.97, 8.73, 10.56    \\
       $20-21$   & 1003  & 0.46 & 2.03, 3 & -0.75 & 8.91, 14.37    \\
       $22-23$   & 1002  & 0.46 & 1.35 & -1, -2 & 5.73, 5.89    \\
  \end{tabular}
  \caption{Overview of the conducted simulations with the associated parameter values. In runs 1-12, the copepods swim in the up-left direction and the initial body orientation angle is 2.36 rad with respect to the horizontal direction. For runs 13-23, the swimming is in the down-right direction and the body orientation angle is -0.65 rad.}
  \label{Tab.Osc}
  \end{center}
\end{table}

\begin{figure}
    \centering
    \includegraphics[width = 1\textwidth]{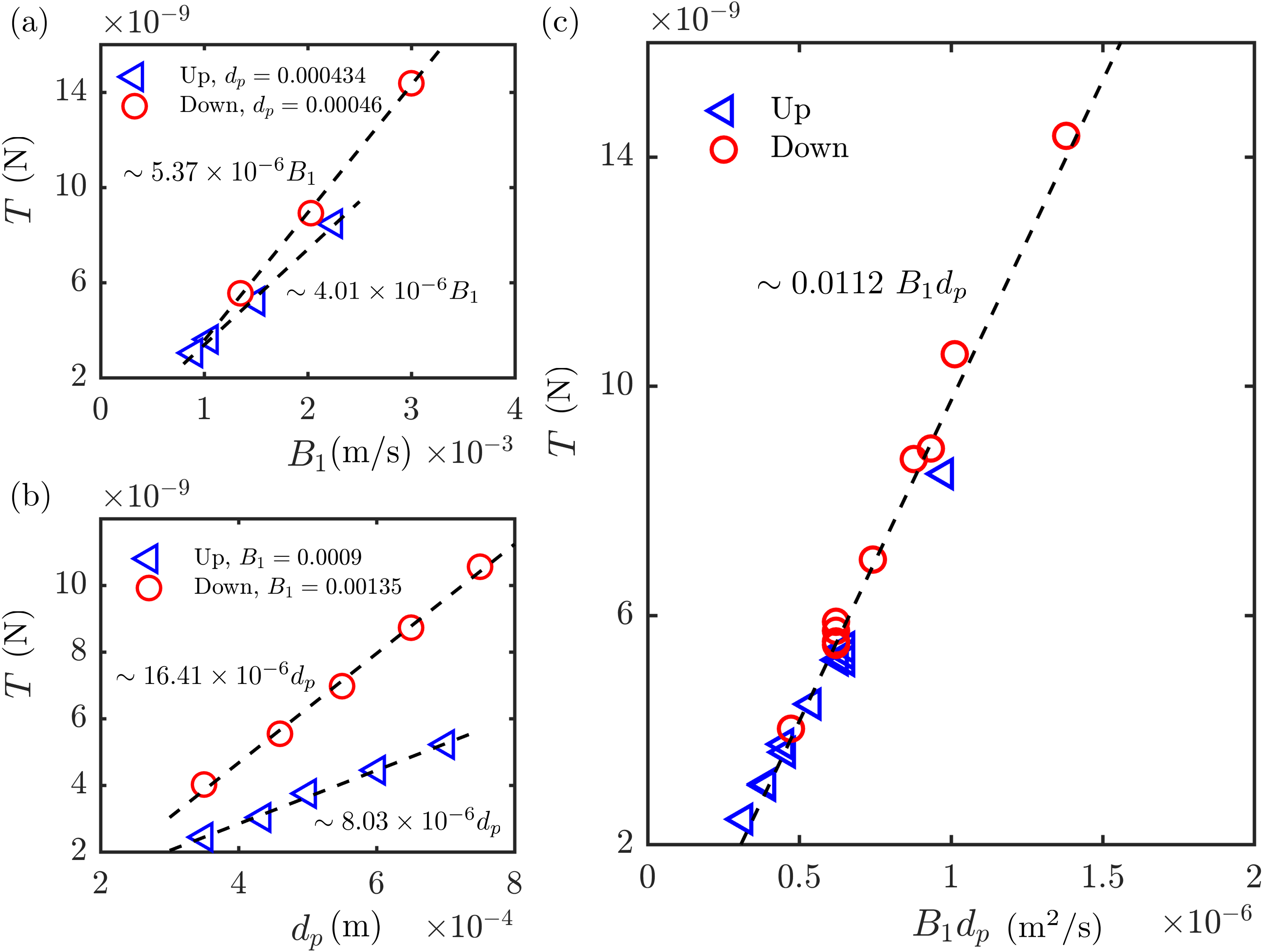}
    \caption{Thrust force evaluation. (a) The thrust force $T$ as a function of the squirmer model parameter $B_1$ for upward and downward swimming directions. The dashed straight lines represent linear fits. (b) The thrust force $T$ as a function of the swimmer size $d_p$. (c) The thrust force $T$ as a function of the expression $B_1d_p$.}
        \label{fig: force analysis}
\end{figure}

Our results demonstrate that the size of a swimmer $d_p$ and its tangential surface velocity field (as imposed in the model via the parameter $B_1$) influence how much thrust is generated, regardless of swimming direction (compare among runs 4-7, 16-19, 8-10, and 20-21). In contrast, the thrust force is insensitive to whether a swimmer is a pusher or a puller and slight variations in its density  (compare among runs 1-3, 13-15, 11-12, and 22-23). Analyzing $T$ as a function of $B_1$ and $d_p$, it becomes evident that the simulation data is well-characterized by linear fits with distinct slopes for upward and downward swimming cases (Fig. \ref{fig: force analysis}(a) and (b)). The linear relationship is also observed in Fig. \ref{fig: force analysis}(c), which effectively collapses all results from Table \ref{Tab.Osc} showing that T scales linearly with the product $B_1d_p$, regardless of swimming direction. This scaling behavior is consistent with the underlying mechanics of the squirmer model, which defines the fluid velocity at discrete Lagrangian marker points distributed over the surface of the swimmer. An increase in $B_1$ results in a proportional enhancement of the slip velocity at these points. Consequently, this leads to an increase in the momentum flux and thus the thrust. On the other hand, an increase in $d_p$ results in an expansion of the area over which active stresses are integrated. Consequently, the total thrust is determined by the combination of the local stress magnitude (dictated by $B_1$) and the total surface area available for force transmission (dictated by $d_p$). Although the available surface area scales as $d^2_p$, the hydrodynamic stress in a Newtonian fluid scales as $1/d_p$ due to the velocity gradient, so that the resulting thrust scales linearly with $d_p$.

 \subsubsection{Swimming in stratified ambient fluid}
To analyze the effect of fluid stratification on BHT, we performed additional simulations of swimmers descending in fluids with different buoyancy frequencies $N=(\gamma g/\rho_0)^{1/2}$, in which $\gamma=\partial \rho/\partial y$ is the density gradient. We mimic salt-stratified fluids by using a Schmidt number $Sc=\nu_f/\kappa_f=1000$ \citep{doostmohammadi2014numerical}. It must be noted that while typical values of $N$ in oceanic pycnoclines are around 0.1 s$^{-1}$, we included higher values to compare against laboratory studies, which are performed under highly stable conditions. To compare active to passive swimming, we also considered settling particles with $B_1 = 0$ and $\beta = 0$ (runs S5-S8) along with neutrally buoyant swimmers, whereby we set the density of the particles to match the density of the ambient fluid at its instantaneous location (runs S9-S12). This setup allows us to isolate the effect of stratification on the motion of a particle or swimmer \citep{more2020motion}. The parameters used in the simulations are listed in Table \ref{Tab.Oss}.

\begin{table}
  \begin{center}
\def~{\hphantom{0}}
  \begin{tabular}{lccccccc}
       Run & $\rho_p\ \mathrm{(kg\ m^{-3})}$ & $d_p\ \mathrm{(m)}$ & $B_1\ \mathrm{(m\ s^{-1})}$ & $\beta$ & $N\ \mathrm{(s^{-1})}$ & $Fr$ & Direction\\[3pt]
       S1-S4   & 1003    & 0.00056  & 0.0015 & -1 & 0.05, 0.1, 0.25, 0.5 & 105, 52.7, 21.5, 10.5 & Downward   \\
       S5-S8   & 1003    & 0.00056  & 0 & 0 & 0.05, 0.1, 0.25, 0.5 & 35.7, 17.9, 7.29, 3.57 & Downward   \\
       S9-S12   & -       & 0.00056  & 0.0015 & -1 & 0.05, 0.1, 0.25, 0.5 & 69.7, 34.9, 14.2, 6.97 & Downward   \\
       S13   & 1003    & 0.00056  & 0.0015 & 1  & 0.50 & 10.5 & Downward    \\
       S14   & 1003    & 0.00056  & 0.0015 & -1  & 0.50 & 10.5 & Upward   \\
       S15   & 1003    & 0.00056  & 0.0015 & 1  & 0.50 & 10.5 & Upward    \\
  \end{tabular}
  \caption{Parameters used for the simulations considering stratified ambient fluid.}
  \label{Tab.Oss}
  \end{center}
\end{table}

\begin{figure}
    \centering
    \includegraphics[width = 0.99\textwidth]{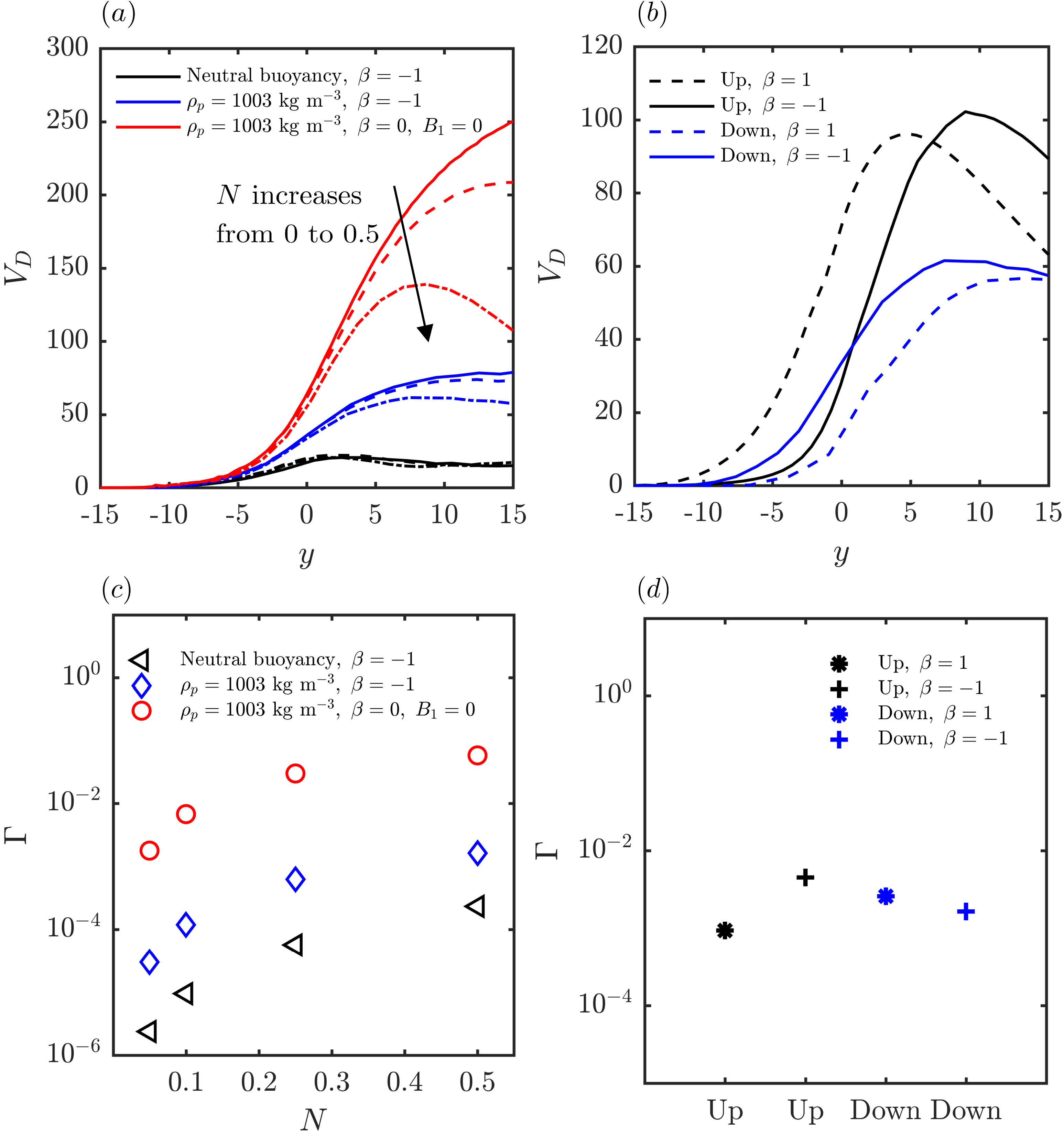}
    \caption{Drift volume in stratified fluid. (a) Drift volume evolution at different density stratification strengths: $N$ = 0 s$^{-1}$ (uniform fluid, solid line), 0.25 s$^{-1}$ (dashed line), and 0.5 s$^{-1}$ (dashdotted line). Other simulation parameters are the same as in Fig.~\ref{fig: drift volume}(d). (b) Drift volume evolution considering vertical upward and downward swimming directions (black and blue lines, respectively) for pushers ($\beta = -1$) and pullers ($\beta$ = 1). The buoyancy frequency in these tests was set at  $N$ = 0.5 s$^{-1}$ (c) The mixing efficiency $\Gamma$ for a settling particle and for active swimmer moving in homogeneous and stratified fluids. (d) The mixing efficiency $\Gamma$ for the pusher and puller cases swimming along different directions. The buoyancy frequency in these tests was set at  $N$ = 0.5 s$^{-1}$.}
        \label{fig: drift volume in stratification}
\end{figure}

For settling particles, the drift volume $V_D$ decreases significantly as the density stratification increases (red lines in Fig. \ref{fig: drift volume in stratification}(a)). In particular, when $N=0.5\ \rm{s}^{-1}$, the drift volume drops after reaching its maximum value due to a pronounced reversed jet formed in the wake of the sinking particle that effectively transports the displaced fluid backward to its equilibrium position. A similar trend was also observed by \citet{doostmohammadi2014numerical}, who demonstrated that displaced isopycnals are forced toward their neutrally buoyant levels by buoyancy effects. For swimmers denser than the ambient fluid (blue lines), the reduction in $V_D$ is less severe than for passive particles with identical excess weight. For neutrally buoyant swimmers (black lines), the drift volume remains independent of the stratification strength. It is important to recognize, however, that the assumption of constant neutral buoyancy in a stratified column requires swimmers to actively regulate their density to compensate for the varying buoyancy forces encountered during vertical motion.

Under strong density stratification ($N=0.5\ \rm{s}^{-1}$), the evolution of the $V_D$ exhibits a pronounced dependence on swimming mode and direction of motion (Fig. \ref{fig: drift volume in stratification}(b)). Compared to cases in homogeneous ambient fluid (Fig. \ref{fig: drift volume}(h)), the drift volume of ascending swimmers undergoes a significantly larger reduction than for their descending counterparts. This intensified reduction is driven by the formation of a more vigorous, stratification-induced reversed jet in the wake of ascending organisms, which facilitates transport back to equilibrium positions (see also \cite{young2010effect}). Interestingly, the upward swimming puller ($\beta = 1$, black dashed line in Fig.~\ref{fig: drift volume in stratification}(b)) is more sensitive to the density stratification than the upward pusher ($\beta = -1$, black solid line in Fig.~\ref{fig: drift volume in stratification}(b)). This heightened sensitivity is rooted in the distinct flow topologies resolved in Fig.~\ref{fig: drift volume} (e) and (f). While pullers entrain fluid from upstream local regions, pushers draw fluid primarily from their lateral sides. Consequently, the upward puller generates a steeper vertical displacement of the density interface, triggering a stronger upward return jet that further diminishes net transport.

In addition to evaluating transport by analyzing the drift volume, we quantified the mixing efficiency $\Gamma$ of the simulated swimmer cases. Following common formulations in the literature \citep{wang2015biogenic,more2020motion}, the kinetic energy equation in a quasi steady state can be written as
 \begin{equation}
     \oint_{S}(\boldsymbol u \cdot \sigma)\cdot \boldsymbol{n} \mathrm{d}S = \frac{2}{Re}\int_{\Omega-\Omega_S}s_{ij}s_{ij}d\Omega - Ri\int_{\Omega-\Omega_S}v\rho' d\Omega,
     \label{equ.KEE}
\end{equation}
where $S$ is the swimmer's surface, $\boldsymbol{n}$ is the unit vector normal to the surface $S$, $\rho'(x,y,z,t)=\rho_f(x,y,z,t)-\rho_f(x,y,z,0)$ is the change of density compared to the initial linear background density, $\Omega$ is the whole fluid domain, $\Omega_s$ is the swimmer domain, and $s_{ij}$ is the rate-of-strain tensor, namely
  \begin{equation}
     s_{ij}=\frac{1}{2}\bigg(\frac{\partial u_i}{\partial x_j}+\frac{\partial u_j}{\partial x_i}\bigg).
     \label{sij} 
 \end{equation}
As introduced in section \ref{sec.Ndf}, the dimesionless parameters $Re=U_{comp}R_p/\nu_f$ and $Ri=gC\alpha R_p/U^2_{comp}$. The term on the left hand side of Eq. \ref{equ.KEE} is the total energy input generated by the swimmer. The first and second terms on the right hand side are the viscous dissipation and the rate of creation of gravitational potential energy in the entire fluid domain, respectively. We define $\Gamma$ as the ratio of the rate of change in background potential energy to the total rate of energy input:
 \begin{equation}
     \Gamma = \frac{-Ri\int_{\Omega-\Omega_S}v\rho'd\Omega}{\frac{2}{Re}\int_{\Omega-\Omega_S}s_{ij}s_{ij}d\Omega-Ri\int_{\Omega-\Omega_S}v\rho' d\Omega}.
     \label{Gamma} 
 \end{equation}
In most of our simulations, the swimming speed changes with time because of the competition between buoyancy and gravity. We evaluate the mixing efficiency for each simulation case by calculating the time-averaged value when a swimmer moves 30 length units in the domain. 
 
Increasing the strength of the density stratification results in more efficient mixing of the surrounding fluid as active and passive particles descend (Fig. \ref{fig: drift volume in stratification}(c)). This is due to the fact that a stronger stratification suppresses vertical Reynolds stresses and turbulent kinetic energy production, thereby leading to a reduction in viscous dissipation. Similar results were reported using numerical simulations of neutrally buoyant swimmers \citep{wang2015biogenic,more2020motion}. In particular, for a given density stratification the settling particles (red circles) induce a higher mixing efficiency than the active swimmers that share the same excess weight (blue diamonds), while the neutrally buoyant swimmers (black triangles) generate the lowest mixing efficiency among the three. This is in line with our results on transport, whereby the vectorial orientation of the thrust force and the co-alignment with the excess weight of the self-propelled swimmer have an important effect on the amount of induced mixing.

We also analyzed the role of swimming mode on BHT by comparing the efficiency of pushers ($\beta=-1$) and pullers ($\beta=1$) ascending and descending within a stable stratification ($N=0.5\ \rm{s}^{-1}$, Fig.~\ref{fig: drift volume in stratification}(d)). For upward-swimming cases, pushers exhibit a higher mixing efficiency than pullers. In contrast, in the case of downward swimmers, pullers mix the fluid marginally more efficiently than pushers. These differences are primarily driven by the spatial extent of the recirculation regions near the body of the swimmers, which have been identified as the primary sites for gravitational potential energy generation \citep{more2020motion}. The near-field flow field and the external gravitational force ultimately determine the scale of these regions: during ascent, the force balance favors the expansion of the recirculation zone for pushers, while during descent, pullers develop a larger recirculatory footprint. This, in turn, facilitates a larger vertical displacement of isopycnals and a subsequent increase in potential energy gain, thereby enhancing the overall mixing efficiency. The difference in mixing efficiency between pushers and pullers for upward swimmers mirrors the trends observed in the drift volume analysis (Fig.~\ref{fig: drift volume in stratification}(b)). In contrast, for downward swimmers, the mixing efficiency exhibits the opposite trend to that of the drift volume. This is because, for upward swimmers, both the drift volume and the mixing efficiency are dominated by recirculating regions, whereas for downward swimmers the drift volume is mainly governed by body-induced drift rather than recirculation. 

 \begin{figure}
    \centering
    \includegraphics[width = 1\textwidth]{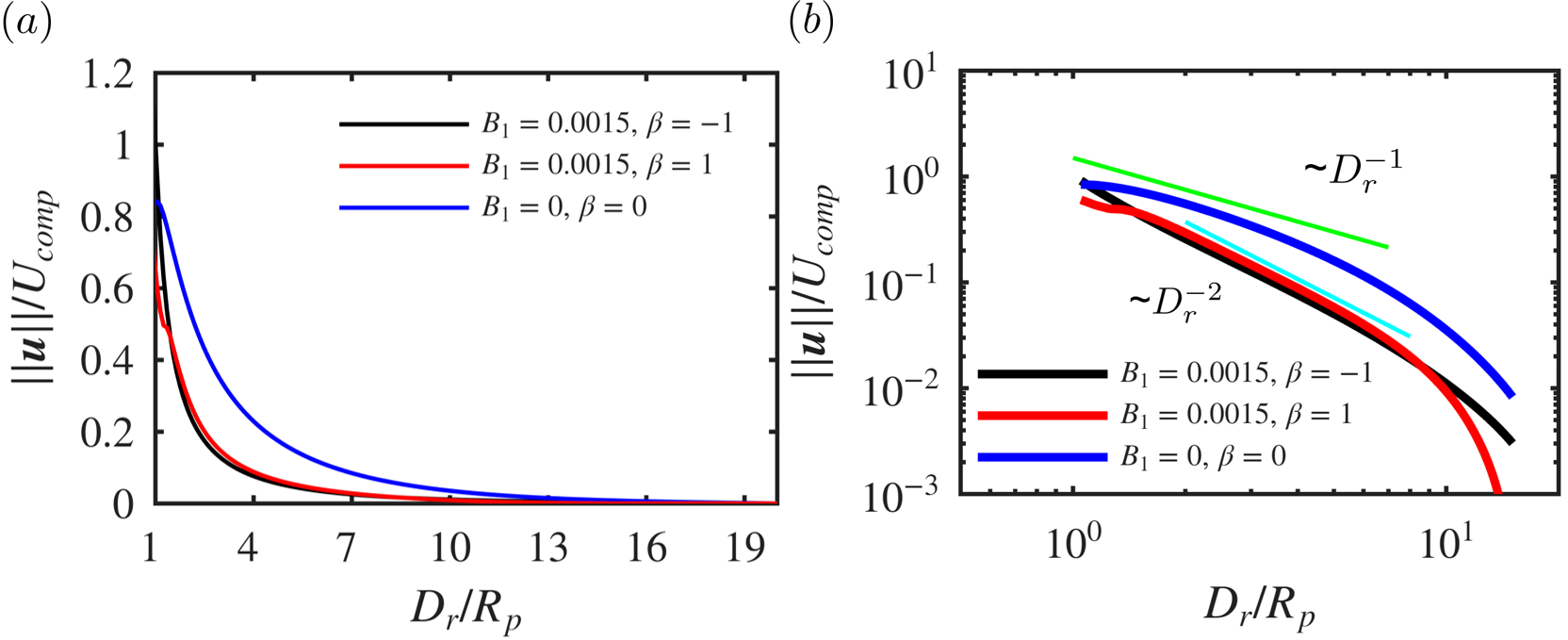}
    \caption{(a) The decay of fluid velocity $||\boldsymbol u||$ along the vertical direction through the center of the swimmers, obtained by averaging the upstream and downstream values. $D_r$ is the distance from the center of the swimmers. (b) The decay of fluid velocity in log-log scale. The solid lines show different slopes.}
        \label{fig: velocity decay}
\end{figure}

 Finally, while subtle differences exist between pushers and pullers during vertical migration,  passive particles consistently achieve significantly higher efficiencies. This divergence stems from the rapid spatial attenuation of the near-field flow generated by self-propelled swimmers relative to passive particles (Fig.~\ref{fig: velocity decay}(a)). In particular, the velocity magnitude $||\boldsymbol u||$ decreases slowly for passive particles compared to self-propelled swimmers. This results in a more extensive isopycnal displacement, and, consequently, greater potential energy generation. This finding underscores a fundamental hydrodynamic trade-off: while active propulsion minimizes fluid disturbances to reduce predator detection---a strategy known as hydrodynamic quietness \cite{kiorboe2014flow}---it simultaneously limits the capacity for efficient BHT. In the present study, the near-field velocity attenuation follows power laws ranging from r$^{-2}$  to r$^{-1}$, consistent with the experimental measurements of \citet{drescher2010direct} and \citet{kiorboe2014flow} (Fig. \ref{fig: velocity decay}(b)).  We further observe that for active swimmers, the near-field velocity decay is markedly steeper than in the far-field; the velocity magnitude drops by 75\% within approximately two radii of the center of the body, whereas a passive particle experiences only a 33\% reduction over the same distance (Fig. \ref{fig: velocity decay}(a)).
 
\section{Conclusions}
In this study, we investigated the near-field flow of free-swimming copepods by using a combined experimental and numerical approach. By focusing on organisms with an inherent excess weight, we characterized how the orientation relative to gravity influences the near-body flow topology and the resulting fluid transport and mixing efficiency. Our experimental results demonstrate an interesting asymmetry, in which organisms achieve higher speeds during descent. We attribute this to the force balance between thrust, drag, and the negative buoyancy of the copepods during vertical migrations. Our PIV measurements illustrate that near-body flow fields also show differences whereby the downward flow generated by the beating appendages is more pronounced during ascent, likely due to the need to overcome gravitational forces.

Through numerical squirmer simulations, we quantified the effect of the observed near-field flow dynamics on BHT in homogeneous and stratified environments via drift volume analysis and evaluation of mixing efficiency, respectively. We observed the so-called tethered-like effect in negatively buoyant swimmers. Their excess weight increases the fore-aft asymmetry of their near-body flow field and, consequently, increases the induced drift volume relative to neutrally buoyant particles. However, stable density stratification effectively suppresses BHT by triggering buoyancy-driven return flows, a phenomenon to which passive settling particles are particularly sensitive. In particular, we found that the upward-swimming pushers generate a much higher mixing efficiency than pullers (approximately five times higher), while the downward-swimming pullers generate a slightly higher mixing efficiency than pushers. However, our observations show that copepods tend to be pullers in upward swimming and pushers in downward swimming, both corresponding to the lower mixing efficiency scenarios. This suggests that copepods tend to adopt a swimming strategy that produces a discrete hydrodynamic signature and, as a result, is less efficient in BHT generation. In addition, we examined the propulsion force generated by the swimming squirmer model and identified that the thrust force exhibited a linear relationship with the product of the particle size $d_p$ and the swimmer strength $B_1$. 

Our future research will focus on the direction effects at the swarm scale. Specifically, we aim to delve deeper into the question of whether the differences in swimming speed and fluid transport exhibited by individual swimmers could accumulate and become significant in the BHT produced during collective migration. In addition, in density-stratified fluids, there is another asymmetry between upward and downward traveling vortices due to the production of baroclinic vorticity. This asymmetry can potentially play an important role in the BHT induced by upward and downward migrating copepod swarms.  We anticipate that the cumulative effects of these small individual swimmers and the asymmetry of vortex dynamics in stratified fluids could potentially have far-reaching effects on the transport of essential nutrients and the overall biogeochemical processes in the ocean ecosystem.

\section{Acknowledgements}
The authors would like to gratefully acknowledge Sara Santos, Nils Tack, and Marjorie Bradley, for their insights and support throughout this work. We also want to thank Andrew Rhyne, Allex Gourlay, Bradford Bourque, and Sean Colin from Roger Williams University for providing the copepods. The authors would also like to thank the NASA Ocean Biology and Biogeochemistry program for providing the financial support for this project (80NSSC22K0284).

Author contributions: Y.S. designed and conducted the experiments, led the analysis of experimental data, contributed to the analysis of simulation results, and drafted the manuscript. R.Z. carried out the numerical simulations, led the analysis of simulation data, and wrote the corresponding sections of the manuscript. E.M. provided project supervision, contributed to the simulation and data interpretation, reviewed and edited the manuscript, and secured project funding. M.M.W. supervised the overall project, contributed to the experimental design and data interpretation, reviewed and edited the manuscript, and secured project funding.

\bibliography{bibliography.bib}
\end{document}